\documentclass[12pt,preprint]{aastex}
\usepackage{enumerate}
\usepackage{float}
\usepackage{placeins}
\usepackage{amsmath}
\usepackage{natbib}
\usepackage{epstopdf}
\usepackage{lscape}
\usepackage{ctable} 
\usepackage{color}
\usepackage[utf8x]{inputenx}
\usepackage[english]{babel}
\usepackage{footnote}
\usepackage{pdflscape}
\usepackage{verbatim}
\usepackage{rotating}
\newcommand{\HI}{H\,{\sc i}~}
\newcommand{\HII}{H\,{\sc ii}~}

\begin{document}

\title{Metal-poor dwarf galaxies in the SIGRID galaxy sample. I. \HII region observations and chemical abundances}
\shorttitle{Low metallicity SIGRID galaxies}
\shortauthors{Nicholls et~al.}

\author{ David C. Nicholls\altaffilmark{1}, Michael A. Dopita\altaffilmark{1}$^,$\altaffilmark{2}, Ralph S. Sutherland\altaffilmark{1}, Helmut Jerjen\altaffilmark{1}, Lisa J. Kewley\altaffilmark{1}, \& Hassan Basurah\altaffilmark{2},}
\email{David.Nicholls@anu.edu.au}
\altaffiltext{1}{Research School of Astronomy and Astrophysics, Australian National University, Cotter Rd., Weston ACT 2611, Australia }
\altaffiltext{2}{Astronomy Department, King Abdulaziz University, P.O. Box 80203, Jeddah, Saudi Arabia}

\begin{abstract}
In this paper we present the results of observations of seventeen \HII regions in thirteen galaxies from the SIGRID sample of isolated gas rich irregular dwarf galaxies. The spectra of all but one of the galaxies exhibit the auroral [O {\sc iii}] 4363\AA~ line, from which we calculate the electron temperature, T$_e$, and gas-phase oxygen abundance.  Five of the objects are blue compact dwarf (BCD) galaxies, of which four have not previously been analysed spectroscopically. We include one unusual galaxy which exhibits no evidence of the [N {\sc ii}] $\lambda\lambda$ 6548,6584\AA~ lines, suggesting a particularly low metallicity  ($<$ Z$_\odot$/30). We compare the electron temperature based  abundances with those derived using eight of the new strong line diagnostics presented by \cite{2013ApJS..208...10D}.  Using a method derived from first principles for calculating total oxygen abundance, we show that the discrepancy between the T$_e$-based and strong line gas-phase abundances have now been reduced to within $\sim$0.07 dex. The chemical abundances are consistent with what is  expected from the luminosity--metallicity relation. We derive estimates of the electron densities and find them to be between $\sim$5 and $\sim$100 cm$^{-3}$. We find no evidence for a nitrogen plateau for objects in this sample with metallicities 0.5~$>$ ~Z$_\odot$~$>$~0.15.

\end{abstract}


\section{Introduction}

The metallicity of \HII regions in  small isolated dwarf galaxies is key to investigating the physical processes that govern star formation in undisturbed stellar systems.\footnote{In this paper we attempt to be explicit in our terminology, using the term ``oxygen abundance'', and referring to ``metallicity'' only in widely used terms such as ``mass--metallicity'' and to refer to total chemical abundances. In addition, the abundance of oxygen measured from spectra is the gas-phase abundance, and does not take into account the oxygen in dust grains.}  
The Small Isolated Gas Rich Irregular Dwarf galaxy (SIGRID) sample of small isolated gas rich irregular dwarf galaxies \citep{2011AJ....142...83N} was selected with the aim of exploring the behavior of the mass-- or luminosity--metallicity relation at the low end of the mass scale. This is based on the observation that nebular metallicity decreases with galaxy stellar mass/luminosity \citep[see, for example,][]{2004ApJ...613..898T, 2006ApJ...647..970L}.  However, the low end of the mass scale shows significantly more scatter in metallicity than the high end in the Tremonti SDSS data. By selecting isolated dwarf galaxies, it was our intention to see if this scatter persisted, and whether it was an intrinsic property of small galaxies. The SIGRID study is complementary to the ``Choirs'' study which looks for tidal dwarf emission line galaxies in group environments \citep{2014ApJ...782...35S}. It is distinct from the Spitzer Local Volume Legacy survey used by \cite{2012ApJ...754...98B} and the SDSS data of \cite{2004ApJ...613..898T} in using targets specifically chosen for their isolation. It is most similar in concept to the study by \cite{2011AstBu..66..255P} of galaxies in the Lynx-Cancer void, but is limited to small very isolated dwarf objects.

Other questions that the SIGRID observations are intended to address are the existence or otherwise of a primary nitrogen ``plateau'' at metallicities below Z=8.45 \citep{1998ApJ...497L...1V}, and the relationship between oxygen abundances determined using ``direct method'', based on the measurement of the electron temperature and the estimation of the ionization correction factors to account for unseen ionization stages, and ``strong line'' technique, based on a calibration of the bright emission lines and emission line ratios.  

There has not been good agreement to date between the two methods, attributed to the empirical nature of the strong line methods.  They have been calibrated in terms of the direct method, and have not until recently had an analytical basis. The direct method has been used as a standard for temperature and metallicity measurement, against which the strong line methods have been calibrated. \cite{2013ApJS..208...10D} subsequently presented a set of strong line diagnostic grids derived from the Mappings photoionization modelling code, based on the latest atomic data \cite[see][]{2013ApJS..207...21N}. We use both the new atomic data and the new diagnostic grids in our analysis.

One might expect there to be greater scatter in the mass--metallicity relation at low masses, due to (1) measurement noise in nebular spectra in fainter galaxies, and (2) different star formation histories in the galaxies. \cite{2006ApJ...647..970L} suggest that the apparent scatter diminishes in observations at longer wavelengths (4.5$\mu$m), and we present additional optical spectral evidence on this question. 

The behavior of the ratio of nitrogen to oxygen abundances at low metallicities also shows increased scatter at lower metallicity.  The current consensus appears to be that there is a low metallicity plateau in log(N/O), indicating the existence of primary nitrogen \citep[see, for example,][]{1993MNRAS.265..199V,1998ApJ...497L...1V,2002MNRAS.330...75C,2006ApJ...636..214V,2010ApJ...720.1738P,2009MNRAS.398..949P,2012ApJ...754...98B,2013ApJ...765..140A}.  However, these previous works were not confined to small isolated dwarf galaxies. Results in our earlier paper on two isolated Local Void dwarf galaxies \citep{2014ApJ...780...88N}, indicated that log(N/O) did not plateau at low metallicity, suggesting no evidence for primary nitrogen.  In this paper we present additional evidence for this.

The paper is structured as follows: in Section 2 we detail the sample selection, the spectroscopic observations, and the data reduction details. H$\alpha$ images of each observed target, spectra, and de-reddened nebular emission line fluxes are presented in Section 3. In Section 4 we present the principal results of the analyses: electron temperatures, gas-phase nebular metallicities with the diagnostic grids, the nitrogen to oxygen flux ratios, the {[}S {\sc ii}{]} line ratios and electron densities, and the luminosity--metallicity results.  In Section 5 we discuss these results, including the anomalies, and in Section 6 we present our conclusions. A discussion of methods used to estimate errors in the emission line fluxes is given in the Appendix.

\section{Observations}
\subsection{Sample selection}
The SIGRID sample was selected to identify small isolated gas-rich irregular dwarf galaxies using the criteria described in \cite{2011AJ....142...83N}. All objects are members of the Survey for Ionization in Neutral Gas Galaxies (SINGG) catalog \citep{2006ApJS..165..307M}, selected from their \HI signatures in the HI Parkes All-Sky Survey (HIPASS) \citep{2004MNRAS.350.1195M} and the presence of H$\alpha$ emission from star forming regions.  From this sample we have, to date, observed 34 objects using IFU optical spectroscopy, as detailed below.  From these observations we report here on 12 galaxies where the [O~{\sc iii}] auroral line is evident, allowing us to calculate the electron temperature T$_e$, and the gas-phase oxygen abundance; and an additional galaxy, J1118-17, with an unusual spectrum with no observed [N~{\sc ii}] lines. In four objects, two separate \HII regions were observed which exhibited the auroral line, resulting in 18 separate \HII region observations. Three  objects (J1152-02, J1225-06, J1328+02) are not members of the final SIGRID sample, but had been observed during the refining of that sample. They were later excluded due to possible influence by regional galaxy groups and clusters, although they are clearly isolated objects, as evidenced by their isolation $\Delta$ index values \citep{2011AJ....142...83N}. Five objects qualify as Blue Compact Dwarf (BCD) galaxies, using the definition by \cite{2002ASPC..273..341S} and discussed by \cite{2011AJ....142...83N}, though they have not previously been identified as such.

\subsection{Spectroscopic observations}

The targets were observed using the WiFeS IFU spectrograph \citep{2007Ap&SS.310..255D, 2010Ap&SS.327...245D} on the Australian National University (ANU) 2.3 m telescope at Siding Spring. The WiFeS instrument is a double-beam image-slicing IFS, designed specifically to maximise throughput. It covers the spectral range 3500--9000\AA, at resolutions of 3000 (full spectral range) and 7000 (long wavelength limit 7000\AA). It has a science field of view (FOV) of 25$\times$38 arc sec. As most of the SIGRID objects subtend angles less than its FOV, WiFeS is an ideal instrument to measure nebular metallicities in the ionized hydrogen star-forming regions. The instrument generates a data cube that allows exploration of nebular and continuum spectra in different regions of the target objects. Typically, even in poor seeing WiFeS resolves SIGRID object star formation regions easily, making possible exploration of excitation and abundances in different regions of each object. In these observations, resolutions used were R=3000 for the blue camera and R=7000 for the red, spanning a usable wavelength range of $\sim$3600 to 7000 \AA. Short period (150 second on object, 75 second on sky) nod-and-shuffle observations were used for all objects, to allow near-complete removal of the sky lines. The exposure times recorded in column 5 of Table \ref{t1} are the on-object integration times.

Details of the observations are given in Table \ref{t1}.  The sample is described in detail in \cite{2011AJ....142...83N}: essentially, the objects lie between redshifts  of 300 and 1650 km/s, have neutral hydrogen masses less than and R-band magnitudes fainter than the Small Magellanic Cloud, low rotation velocities, show evidence of current active star formation, and are isolated, away from galaxy clusters and the tidal effects of other galaxies.  All objects exhibit low (gas-phase) oxygen abundance (log(O/H) $\lesssim$ Z$_{0.3\odot}$), as we describe below.  The seeing listed in column 7 shows that in all but one case, (J1403-27), the seeing was average for Siding Spring. Even in that case, the seeing was better than the spaxel sample size, resulting in little if any flux loss.

Several classes of object were identified in \cite{2011AJ....142...83N}, including ``bloaters'', which are objects considerably more spatially extended than their masses would suggest.  One of these is J1118-17.  It is very faint, but as we show below, appears to have a very low metallicity. In the light of the results obtained for the relatively faint object, J1118-17 (s1 and s2 targets), it would have been desirable to undertake significantly longer integration times, but observing conditions (weather and moonlight) did not permit this.  We intend to undertake further longer integration time observations for this unusual object. The spectrum of J1118-17s1 is very noisy with few usable spectral lines, so we have analysed only s2---the two objects appear similar apart from luminosity. We reported results for J1609-04, a very isolated galaxy at the edge of the Local Void, in a previous paper \citep{2014ApJ...780...88N}, and the results are included again here for completeness.

 \ctable[
caption= {Observations of objects from SIGRID sample},
pos=h,
captionskip=3pt,
sideways,
label=t1,
doinside=\scriptsize,
]{llcccclccccll}
{
\tnote[1]{Object data from \citet{2011AJ....142...83N}}
\tnote[2]{Columns 1,2: object ID; Columns3,4: coordinates; Column 5: observation date; Column 6: exposure time on object; Column 7: regions with [O~{\sc iii}] auroral line; Column 8: distance (Mpc); Column 9: neutral hydrogen mass; Column 10: R-band-magnitude; Column 11: isolation index; Column 12: comments  }
} 
{\toprule[1.5pt]
Object & Alternate ID		&	RA & Decl. & Observed & Exp. & seeing	&aur. & D & log(m$_{HI}$) & M$_R$   & Delta  & Comments \\
(HIPASS ID) &	(NED)	&	 (J2000) & (J2000) & (date) & (min.) & (arc sec) & (\#) & (Mpc) & (m$_{H_\odot}$) & (mag.) & index & \\
\midrule
J0005-28		&	ESO149-G013	&	00 05 31.8	&	-28 05 53	&	27-Aug-11		&	60	&	1.5	&	1	&	10.2	&	8.23	&	-15.3		&	-2.1	&	BCD \\
J1118-17(s2)	&	n/a			&	11 18 03.1		&	-17 38 31	&	13-Mar-11		&	80	&	1.5	&	0	&	13.5	&	8.56	&	-13.5		&	-2.2	&	v. low N~{\sc ii} \\
J1152-02A,B	&	UGC 06850	&	11 52 37.2		&	-02 28 10	&	07-Mar-11		&	60	&	2.2	&	2	&	13.5	&	8.31	&	-16.7		&	-1.7	&	BCD \\
J1225-06s2	&	LEDA 1031551	&	12 25 40.0	&	-06 33 07	&	11-May-10	&	60	&	2.8	&	1	&	20.2	&	8.48	&	-14.2		&	-1.5	&	\\
J1328+02		&	LEDA 135827	&	13 28 12.1	&	+02 16 46	&	13-May-10	&	40	&	1.8	&	1	&	15.5	&	7.93	&	-15.2		&	-0.7	&	\\
J1403-27		&	ESO510-IG052	&	14 03 34.6	&	-27 16 47	&	11-May-10	&	60	&	3-5	&	1	&	17.5	&	8.72	&	-16.6		&	-1.8	&	BCD \\
J1609-04[2][5]	&	MCG-01-41-006&	16 09 36.8	&	-04 37 13	&	25-Aug-11		&	60	&	2-2.4	&	2	&	14.8	&	8.30	&	-16.1		&	-2.9	&	\\
J2039-63A,B	&	LEDA 329372	&	20 38 57.2	&	-63 46 16	&	16-Sep-09		&	60	&	1.3	&	2	&	22.8	&	8.31	&	-16.5		&	-1.4	&	BCD \\
J2234-04B	&	MCG-01-57-015&	22 34 54.7	&	-04 42 04	&	26-Aug-11		&	60	&	1.3	&	1	&	20.5	&	8.50	&	-16.2		&	-0.2	&	\\
J2242-06		&	LEDA 102806	&	22 42 23.5	&	-06 50 10	&	09-Jul-10		&	60	&	1.8-2	&	1	&	14.1	&	7.95	&	-15.6		&	-0.7	&	\\
J2254-26		&	MCG-05-54-004&	22 54 45.2	&	-26 53 25	&	16-Sep-09		&	60	&	1.3	&	1	&	12.1	&	8.46	&	-16.1		&	-2.1	&	BCD \\
J2311-42A,B	&	ESO291-G003	&	23 11 10.9		&	-42 50 51	&	27-Aug-11		&	60	&	1.8-2	&	2	&	19.1	&	8.19	&	-16.5		&	-1.3	&	\\
J2349-22		&	ESO348-G009	&	23 49 23.5	&	-22 32 56	&	06-Oct-10		&	80	&	1.8-2	&	1	&	7.7	&	7.99	&	-14.7		&	-0.5	&	\\
\addlinespace[5pt]\bottomrule[1.5pt]\addlinespace[5pt]}

\FloatBarrier

\subsection{Data reduction}

The data were reduced using the revised WiFeS Python ``Pywifes'' pipeline \citep{2013Ap&SS.tmp..406C}. This involves steps generally similar to those described in \cite{2010Ap&SS.327...245D} for the older pipeline: bias modelling and subtraction, flat fielding, arc line identification and wavelength calibration, cosmic ray removal, sky-line subtraction using nod-and-shuffle, initial data cube construction and atmospheric dispersion correction, standard flux star calibration, telluric correction, assembly into the final data cubes and, where necessary, combination of multiple cubes into a final object data cube. The standard stars used were taken from \cite{1999PASP..111.1426B}. Spectral sampling was undertaken using a 6 arc sec diameter circular spatial area centered on each \HII region, through the full wavelength range of the data cube, to obtain spectra for each region. Line fluxes were measured from these spectra using IRAF/splot.  Particular care was taken to account for any stellar absorption features underlying the Balmer emission lines, although in all cases, this was minor or absent, due to low stellar continuum.  In fact, the stellar continuum was extremely faint, with the exception of the object J0005-28 (see Figure \ref{fig_7}, displayed on a log-intensity scale). Unlike single slit spectra, with IFU data cubes, we are able to select the entire area of the \HII region from which to extract the spectrum, and exclude the majority of the galaxy stellar background, resulting in better signal-to-noise. Test sample sizes showed that all the detectable H-alpha and [OIII] in each HII region lay within the sample aperture, except where there are closely adjacent HII regions (e.g., J1609-04), where limiting the sample size to 6 arc sec diameter avoids sampling a different region. Ideally, single spaxel-based analysis would be preferable to multi-spaxel sampling, but these objects are so faint that the resultant noise is prohibitive.

Flux de-reddening was performed on the raw flux data using two methods.  First, for consistency with other work, we used the dust reddening formulae from \cite{1989ApJ...345..245C} with A$_V$=3.1, using the resultant Balmer line flux ratios as a check.  To confirm these results, we used the dust models from \cite{2005ApJ...619..340F}, using a relative extinction curve with R$^A_V$=4.3, where R$^A_V$ = A$_V$/(E$_{B-V}$) and A$_V$ is the V-band extinction.  This is discussed in more detail in \citet[Appendix 1]{2013ApJ...768..151V}. We used an initial Balmer decrement ratio of 2.82 for H$\alpha$/H$\beta$, corresponding to an electron temperature of 12 500K, adjusted the electron temperature using the direct method derived from the [O~{\sc iii}] line ratios, then adjusted the apparent Balmer ratios by varying the value of A$_V$ for the best fit to the H$\gamma$/H$\beta$ ratio, using the ratio H$\delta$/H$\beta$ as a check, fitting to the \cite{1995MNRAS.272...41S} Case B Balmer ratios. The de-reddened flux values reported in Table \ref{t2}  are those using the Cardelli method. In all cases, the two approaches gave similar results (to within $\sim$3\% in the de-reddened Balmer line ratios). In only one case, J2234-04, object A, did the de-reddening fail to provide a plausible result, and this has been excluded from the results reported here.  It appears likely that two or more incompletely removed cosmic ray artefacts were the cause of the problem.

\section{Results}

\subsection{H$\alpha$ images}

\begin{figure}[htbp]
\centering
\includegraphics[width=0.8\hsize]{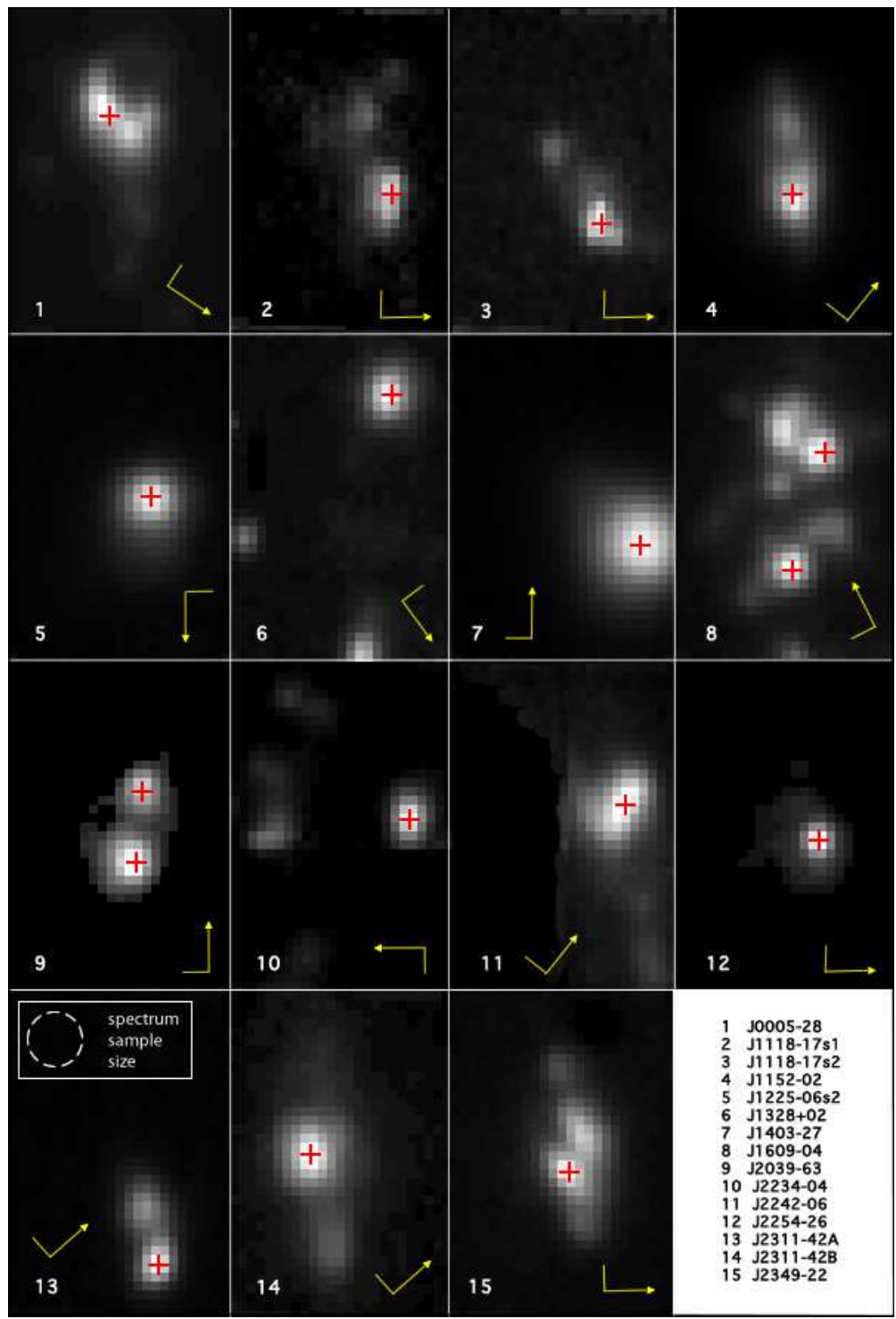}
\caption{H$\alpha$ slices from WiFeS 35$\times$28 arc sec image cubes. The size of the areas sampled to extract spectra is shown in panel 13. North is indicated by the long arrow and East by the bar ($\pm15 ^\circ$). The red crosses mark the center of the sampled areas. Note that the images have been stretched to show the fainter areas.}\label{fig_1}
\end{figure}

Images of the objects listed in Table \ref{t1} are shown in Figure \ref{fig_1}. These are 38$\times$25 arc sec H$\alpha$ slices from the WiFeS data cubes.  Spectra were extracted from these cubes using 6 arc sec diameter samples, centered on each (bright) \HII region. Note that the seeing during the observations of object 7, J1403-27, was poor---3 to 5 arc sec---so the dimensions of the image do not indicate the true size of the \HII region. The image scaling does not reflect the true brightness, but has been adjusted to illustrate the extent of the fainter parts of the  \HII regions.  The sample size is larger than the worst seeing so avoids any sample size flux losses. In all but the one case the sample size is much larger than the object (see Table \ref{t1}).

\subsection{Spectra}
\begin{figure}[htbp]
\centering
\includegraphics[width=0.8\hsize]{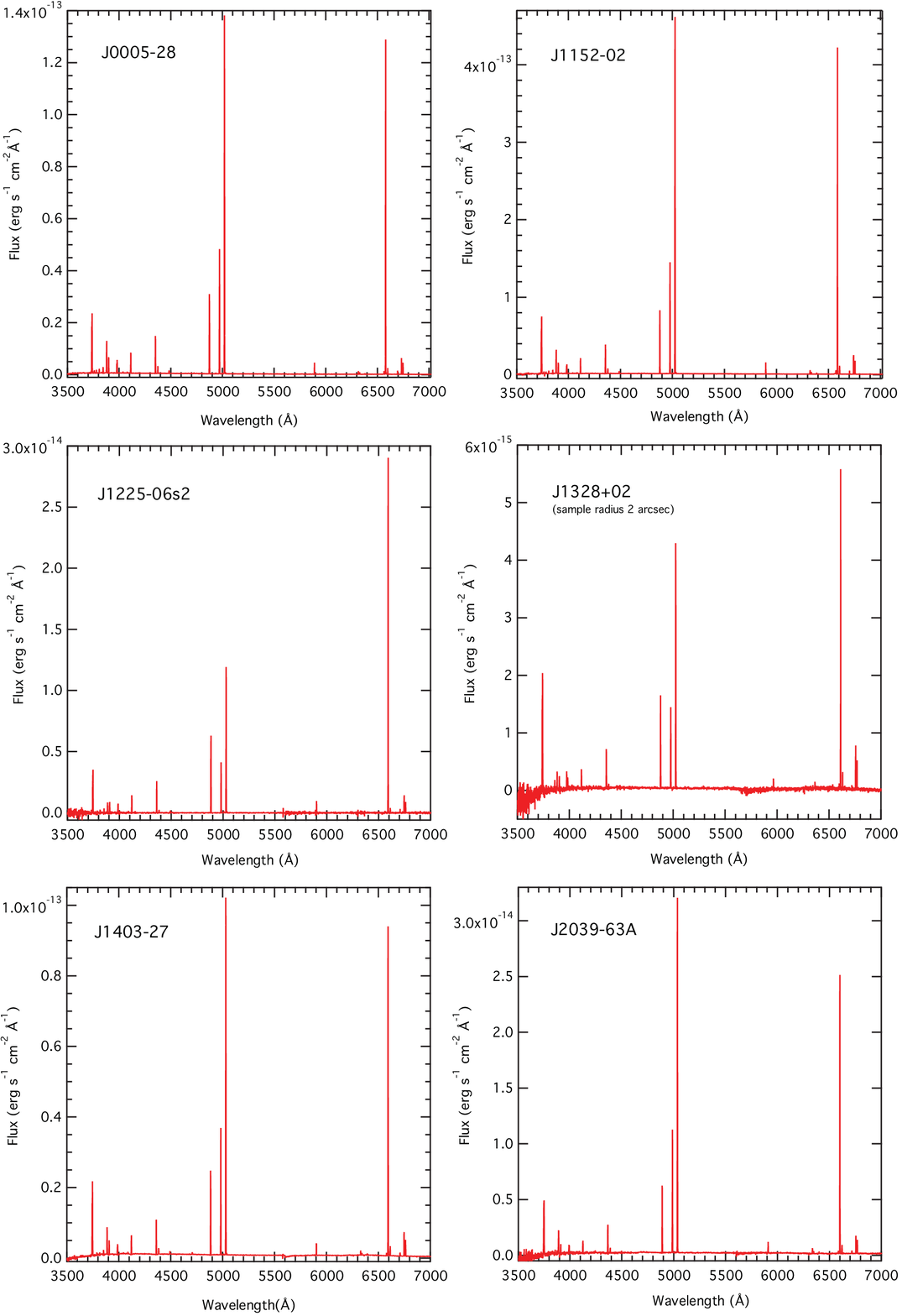}
\caption{Spectra for J0005-28, J1152-02, J1225-06s2, J1328+02, J1403-27, and J2039-63A}\label{fig_2}
\end{figure}

\begin{figure}[htbp]
\centering
\includegraphics[width=0.8\hsize]{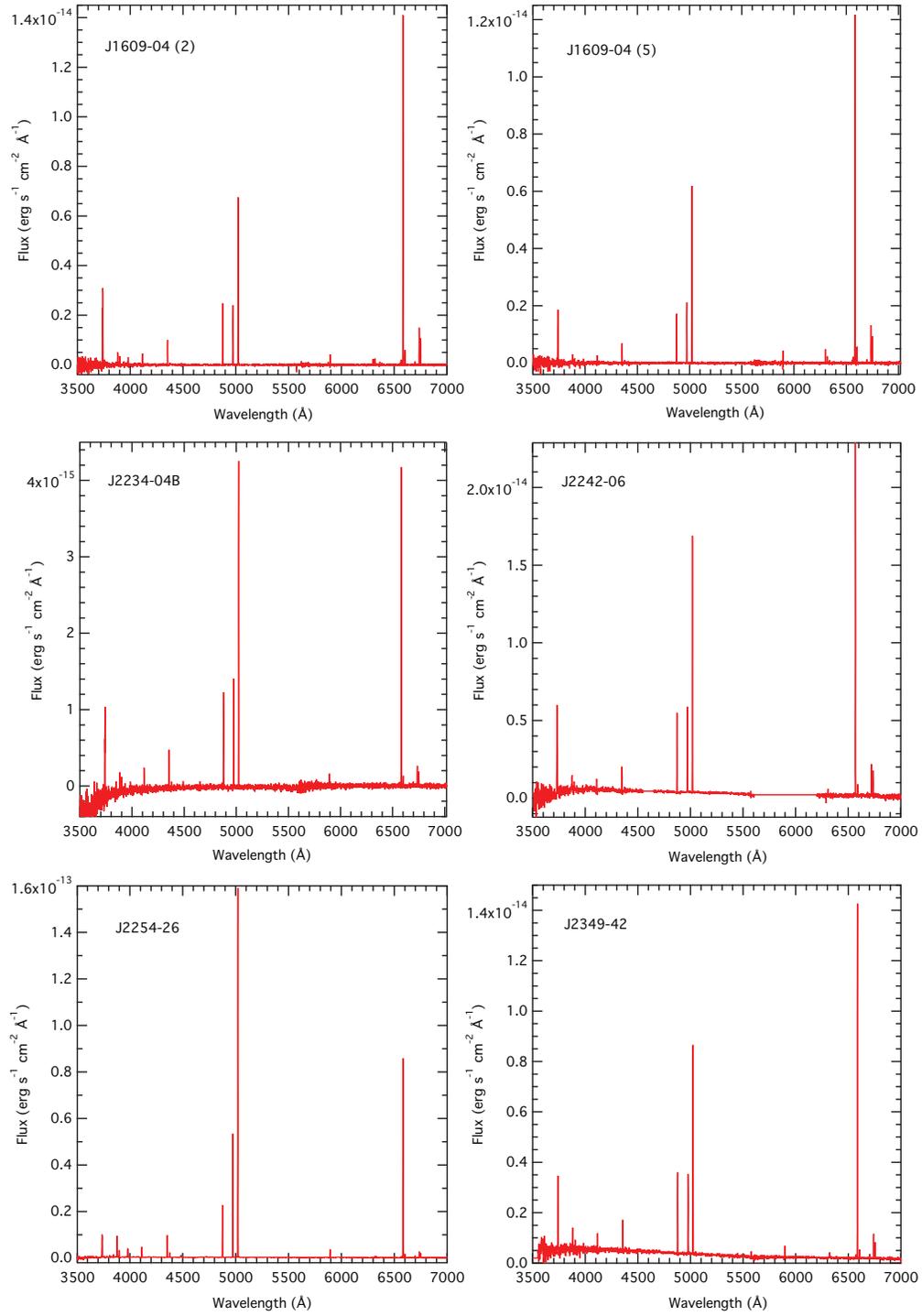}
\caption{Spectra for J1609-04(2), J1609-04(5), J2234-04B, J2242-06, J2254-26, and J2349-42}\label{fig_3}
\end{figure}

\begin{figure}[htbp]
\centering
\includegraphics[width=0.8\hsize]{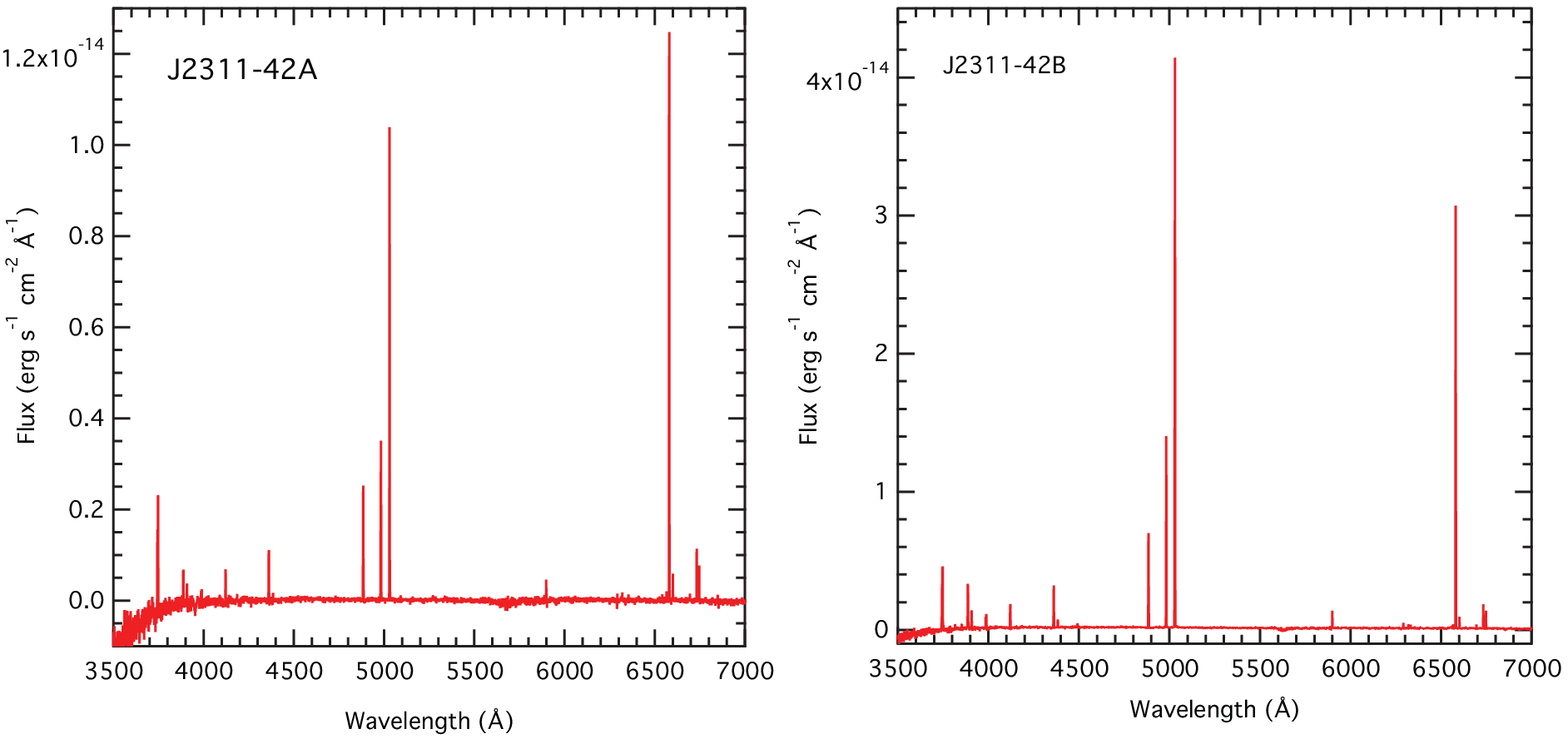}
\caption{Spectra for J2311-42A and J2311-42B}\label{fig_4}
\end{figure}

\begin{figure}[htbp]
\centering
\includegraphics[width=0.65\hsize]{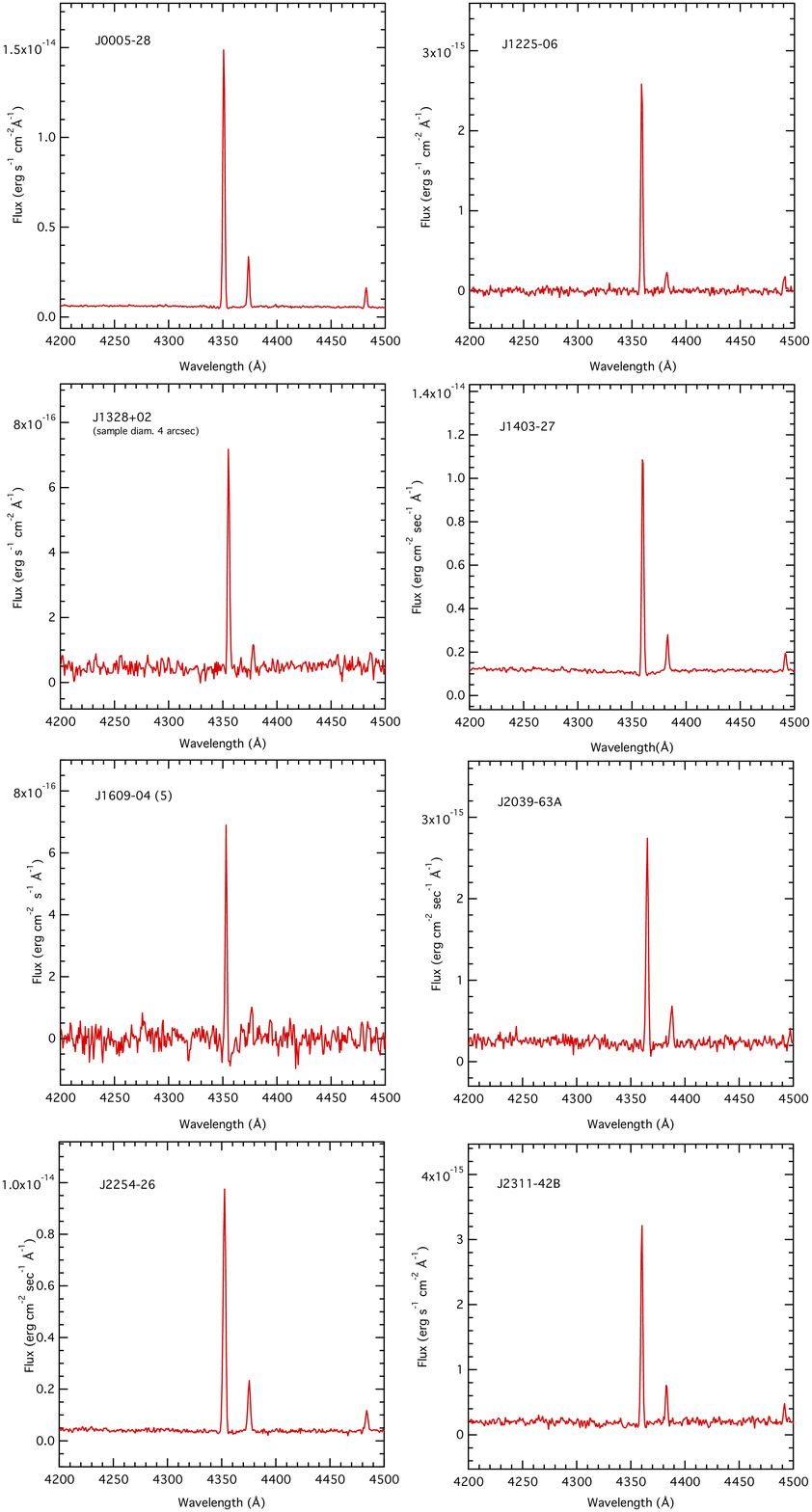}
\caption{Close up of the H$\gamma$ and [O~{\sc iii}] 4363\AA~ lines for eight of the SIGRID objects.}\label{fig_4B}
\end{figure}

Spectra extracted from the WiFeS data cubes are shown in Figures \ref{fig_2}--\ref{fig_6}. The current data reduction pipeline creates a ``sag'' artefact for wavelengths shorter than $\sim$4000 \AA, in the absence of a strong stellar continuum\footnote{This is discussed in detail in \cite{2013Ap&SS.tmp..406C}. It is an artefact of the data reduction pipeline that occurs for objects with very little stellar continuum, observed with the older WiFeS CCD detectors, whereby the spectrum ``droops'' at either end of the passband. It does not affect the flux measurements for individual emission lines. It is most evident in Figure 3, middle left pane.}.  Stellar continua are weak or non-existent in most objects except for J0005-28, J2242-06, and J2349-42.  Defective CCD chip amplification at the time of the observations caused two high noise regions in the spectrum of J2242-06, which have been replaced in Figure \ref{fig_3} with straight lines. Incipient noise from these two chip amplifiers is apparent in other spectra, but does not impact on any of the important diagnostic emission lines. De-reddened fluxes, equivalent widths, and logarithmic extinction coefficients (c(H$\beta$)) for the observed optical nebular lines are shown in Table \ref{t2}. As noted in the appendix, the extinction coefficients were calculated using the Cardelli reddening law with A$_V$=3.1  Similar results ($\pm$3\%) were obtained using methods derived from \cite{2005ApJ...619..340F}. The equivalent widths are large for for some objects, indicating the very low continua, because the host galaxies are very small and faint, and the spectra were measured from an area sampling only the immediate area of the \HII region.

Figure \ref{fig_4B} is a close up of the spectra in Figures \ref{fig_2} to \ref{fig_4}, from 4200\AA~to 4500\AA, illustrating the H$\gamma$ and auroral [O~{\sc iii}] 4363\AA~lines.  The signal to noise is mainly very good, but for three of the 14 objects in Figures \ref{fig_2} to \ref{fig_4}, the detections are real but noisy (see Table \ref{t2}).
\FloatBarrier 

\ctable[
caption= {Reddening corrected line fluxes normalised to H$\beta$=100, with measured H$\beta$ (erg s$^{-1}$ cm$^{-2}$ \AA$^{_-1}$), equivalent widths for H$\alpha$ and H$\beta$, and extinction coefficient, c(H$\beta$). For a discussion of the errors, see appendix.},
pos=h,
captionskip=1pt,
%sideways,
label=t2,
doinside=\tiny,
]{lllllll}
{} 
{\toprule%[1.5pt]
\multicolumn{7}{c}{$I(\lambda)/I(H\beta) \times 100$}              \ML
Ion				&	J0005-28			&	J1118-17s2		&	J1152-02A		&	J1152-02B		&	J1225-06s2		&	J1328+02 \\
\midrule
{[}O~{\sc ii}{]} 3726 	&	 90.39 $\pm$  2.94	&	56.17  $\pm$ 16.48	&	119.11 $\pm$  4.84	&	139.31 $\pm$  6.67	&	 77.69 $\pm$  3.29	&	183.97 $\pm$  7.56 \\
{[}O~{\sc ii}{]} 3729	&	117.91 $\pm$  3.76	&	188.64 $\pm$ 20.45	&	171.01 $\pm$  6.40	&	200.03 $\pm$  8.49	&	108.84 $\pm$  4.23	&	240.15 $\pm$  9.24 \\
{[}Ne~{\sc iii}{]} 3869	&	 39.98 $\pm$  2.31	&	---				&	 49.23 $\pm$  1.48	&	 46.29 $\pm$  1.64	&	 16.37 $\pm$  0.96	&	 20.30 $\pm$  1.47 \\
H$\delta$ 4102 	&	 26.35 $\pm$  0.86	&	28.45  $\pm$ 6.61	&	 25.91 $\pm$  0.84	&	 26.31 $\pm$  0.87	&	 26.26 $\pm$  1.13	&	 23.95 $\pm$  1.28 \\
H$\gamma$ 4340	&	 47.30 $\pm$  1.47	&	35.79  $\pm$ 5.61	&	 41.11 $\pm$  1.29	&	 42.75 $\pm$  1.35	&	 48.01 $\pm$  1.61	&	 46.40 $\pm$  1.86 \\
{[}O~{\sc iii}{]} 4363	&	  9.11 $\pm$  0.32	&	---				&	  6.82 $\pm$  0.25	&	  6.37 $\pm$  0.26	&	  4.83 $\pm$  0.35	&	  5.24 $\pm$  0.49 \\
H$\beta$ 4861		&	100.00 $\pm$  3.03	&	100.00 $\pm$ 5.28	&	100.00 $\pm$  3.03	&	100.00 $\pm$  3.05	&	100.00 $\pm$  3.21	&	100.00 $\pm$  3.30  \\
{[}O~{\sc iii}{]} 4959	&	150.49 $\pm$  4.54	&	20.84  $\pm$ 2.67	&	172.31 $\pm$  5.20	&	146.69 $\pm$  4.46	&	 62.66 $\pm$  2.02	&	 83.86 $\pm$  2.89 \\
{[}O~{\sc iii}{]} 5007	&	451.97 $\pm$ 13.58	&	79.06  $\pm$ 4.51	&	517.38 $\pm$ 15.55	&	439.80 $\pm$ 13.24	&	190.32 $\pm$  6.11	&	256.14 $\pm$  8.06  \\
{[}O~{\sc i}{]} 6300	&	  2.69 $\pm$  0.13	&	---				&	  3.55 $\pm$  0.13	&	  3.73 $\pm$  0.15	&	  1.87 $\pm$  0.41	&	  ---	\\
{[}S~{\sc iii}{]} 6312	&	  1.62 $\pm$  0.08	&	---				&	  1.73 $\pm$  0.07	&	  1.76 $\pm$  0.09	&	---				&	  ---	\\
{[}N~{\sc ii}{]} 6548	&	  2.31 $\pm$  0.11	&	---				&	  2.25 $\pm$  0.09	&	  2.19 $\pm$  0.10	&	---				&	  3.36 $\pm$  0.56 \\
H$\alpha$ 6563	&	279.20 $\pm$  8.41	&	282.90 $\pm$ 10.21	&	282.48 $\pm$  8.63	&	281.77 $\pm$  8.50	&	277.53 $\pm$  8.58	&	279.03 $\pm$  8.87  \\
{[}N~{\sc ii}{]} 6584	&	  4.76 $\pm$  0.17	&	1.48   $\pm$ 1.63	&	  6.99 $\pm$  0.25	&	  6.54 $\pm$  0.25	&	  4.05 $\pm$  0.35	&	 14.65 $\pm$  0.94 \\
{[}S~{\sc ii}{]} 6716	&	 11.92 $\pm$  0.38	&	11.72  $\pm$ 1.82	&	 15.64 $\pm$  0.57	&	 17.38 $\pm$  0.58	&	 11.85 $\pm$  0.61	&	 33.53 $\pm$  1.41 \\
{[}S~{\sc ii}{]} 6731	&	  8.76 $\pm$  0.29	&	6.98   $\pm$ 1.67	&	 11.29 $\pm$  0.40	&	 12.38 $\pm$  0.43	&	  8.27 $\pm$  0.44	&	 21.78 $\pm$  1.11 \\
\midrule
H$\beta$ 4861		&	8.52e-14			&	1.16E-15			&	2.90E-13			&	1.34E-13			&	2.10E-14			&	3.56E-15 \\
EW(H$\alpha$)		&	685				&	717				&	1032				&	642				&	2482				&	234		\\
EW(H$\beta$)		&	187				&	---				&	122				&	77.3				&	120				&	118		\\
c(H$\beta$)		&	0.068			&	0.169			&	0.128			&	0.021			&	0.186			&	0.024	\\
\midrule
Ion				&	J1403-27			&	J1609-04(2)		&	J1609-04(5)		&	J2039-63A		&	J2039-63B		&	J2234-04B \ML
{[}O~{\sc ii}{]} 3726 	&	122.74 $\pm$  4.19	&	229.01 $\pm$ 10.05	&	209.89 $\pm$  9.66	&	133.55 $\pm$  5.98	&	 75.86 $\pm$  4.63	&	144.35 $\pm$  7.89 \\
{[}O~{\sc ii}{]} 3729	&	165.52 $\pm$  5.47	&	255.80 $\pm$ 10.86	&	230.35 $\pm$ 10.27	&	182.27 $\pm$  7.44	&	103.61 $\pm$  5.47	&	137.53 $\pm$  7.69 \\
{[}Ne~{\sc iii}{]} 3869	&	 38.81 $\pm$  1.32	&	 25.85 $\pm$  2.04	&	 30.13 $\pm$  3.13	&	 50.07 $\pm$  2.42	&	 38.62 $\pm$ 13.92	&	 27.83 $\pm$ 18.80 \\
H$\delta$ 4102 	&	 27.67 $\pm$  0.97	&	 27.20 $\pm$  1.67	&	 20.43 $\pm$  1.69	&	 25.63 $\pm$  1.25	&	 24.68 $\pm$  1.44	&	 24.67 $\pm$  2.00 \\
H$\gamma$ 4340	&	 48.07 $\pm$  1.56	&	 49.41 $\pm$  2.04	&	 48.30 $\pm$  2.18	&	 49.77 $\pm$  1.85	&	 48.82 $\pm$  2.25	&	 40.75 $\pm$  2.15 \\
{[}O~{\sc iii}{]} 4363	&	  7.66 $\pm$  0.32	&	  2.27 $\pm$  0.39	&	  6.10 $\pm$  1.09	&	  9.72 $\pm$  0.55	&	  8.39 $\pm$  0.91	&	  6.41 $\pm$  0.93 \\
H$\beta$ 4861		&	100.00 $\pm$  3.09	&	100.00 $\pm$  3.47	&	100.00 $\pm$  3.49	&	100.00 $\pm$  3.21	&	100.00 $\pm$  3.27	&	100.00 $\pm$  3.78 \\
{[}O~{\sc iii}{]} 4959	&	142.26 $\pm$  4.33	&	 91.29 $\pm$  3.11	&	110.15 $\pm$  3.77	&	172.74 $\pm$  5.34	&	156.84 $\pm$  5.05	&	118.89 $\pm$  4.31 \\
{[}O~{\sc iii}{]} 5007	&	420.32 $\pm$ 12.66	&	266.71 $\pm$  8.41	&	322.78 $\pm$ 10.14	&	502.15 $\pm$ 15.21	&	462.63 $\pm$ 14.22	&	351.19 $\pm$ 11.35  \\
{[}O~{\sc i}{]} 6300	&	  3.66 $\pm$  0.18	&	  5.66 $\pm$  0.46	&	  6.65 $\pm$  0.56	&	  5.32 $\pm$  0.32	&	  2.79 $\pm$  0.51	&	  ---	\\
{[}S~{\sc iii}{]} 6312	&	  1.97 $\pm$  0.12	&	  0.00 $\pm$  0.00	&	  ---				&	  1.83 $\pm$  0.21	&	  ---				&	  ---	\\
{[}N~{\sc ii}{]} 6548	&	  2.32 $\pm$  0.15	&	  6.18 $\pm$  0.60	&	  4.96 $\pm$  0.51	&	  2.76 $\pm$  0.25	&	  ---				&	  ---	\\
H$\alpha$ 6563	&	280.10 $\pm$  8.52	&	285.58 $\pm$  9.95	&	279.82 $\pm$  8.72	&	279.63 $\pm$  8.55	&	280.15 $\pm$  8.72	&	280.09 $\pm$  9.14 \\
{[}N~{\sc ii}{]} 6584	&	  7.69 $\pm$  0.31	&	 11.92 $\pm$  0.60	&	 12.58 $\pm$  0.79	&	  8.65 $\pm$  0.43	&	  5.69 $\pm$  0.57	&	  7.87 $\pm$  0.85 \\
{[}S~{\sc ii}{]} 6716	&	 19.52 $\pm$  0.69	&	 29.77 $\pm$  1.15	&	 27.18 $\pm$  1.17	&	 19.11 $\pm$  0.75	&	 12.29 $\pm$  0.66	&	 17.05 $\pm$  1.17 \\
{[}S~{\sc ii}{]} 6731	&	 13.87 $\pm$  0.52	&	 20.96 $\pm$  0.89	&	 18.69 $\pm$  0.91	&	 14.22 $\pm$  0.60	&	  8.19 $\pm$  0.68	&	 12.28 $\pm$  1.09 \\
\midrule
H$\beta$ 4861		&	9.07E-14			&	1.57E-14			&	1.72E-14			&	2.52E-14			&	7.05E-15			&	2.39E-15 \\
EW(H$\alpha$)		&	286				&	215				&	157				&	301				&	459				&	---		\\
EW(H$\beta$)		&	66.2				&	77.6				&	48.1				&	69.5				&	150				&	---		\\
c(H$\beta$)		&	0.216			&	0.448			&	0.619			&	0.219			&	0.028			&	0		\\
\midrule
Ion				&	J2242-06			&	J2254-26			&	J2311-42A		&	J2311-42B		&	J2349-22 			& \ML
{[}O~{\sc ii}{]} 3726 	&	132.13 $\pm$  6.17	&	 66.71 $\pm$  2.46	&	137.44 $\pm$  7.49	&	 92.28 $\pm$  3.86	&	113.10 $\pm$  5.82	& \\
{[}O~{\sc ii}{]} 3729	&	180.06 $\pm$  7.60	&	 83.32 $\pm$  2.95	&	206.55 $\pm$  9.57	&	106.01 $\pm$  4.27	&	155.67 $\pm$  7.10	& \\
{[}Ne~{\sc iii}{]} 3869	&	 28.65 $\pm$  1.89	&	 55.79 $\pm$  1.83	&	 36.57 $\pm$  2.56	&	 51.11 $\pm$  2.01	&	 34.21 $\pm$  2.53	& \\
H$\delta$ 4102 	&	 26.87 $\pm$  1.45	&	 25.74 $\pm$  0.87	&	 30.15 $\pm$  1.93	&	 27.72 $\pm$  1.20	&	 22.54 $\pm$  1.79	& \\
H$\gamma$ 4340	&	 44.07 $\pm$  1.76	&	 46.15 $\pm$  1.47	&	 51.15 $\pm$  2.21	&	 47.21 $\pm$  1.70	&	 44.22 $\pm$  2.12	& \\
{[}O~{\sc iii}{]} 4363	&	  5.07 $\pm$  0.63	&	 10.11 $\pm$  0.39	&	  5.61 $\pm$  0.80	&	  9.45 $\pm$  0.54	&	  4.87 $\pm$  0.97	& \\
H$\beta$ 4861		&	100.00 $\pm$  3.26	&	100.00 $\pm$  3.05	&	100.00 $\pm$  3.52	&	100.00 $\pm$  3.25	&	100.00 $\pm$  3.48	& \\
{[}O~{\sc iii}{]} 4959	&	 99.31 $\pm$  3.23	&	224.02 $\pm$  6.75	&	133.40 $\pm$  4.54	&	192.18 $\pm$  5.94	&	 95.81 $\pm$  3.39	& \\
{[}O~{\sc iii}{]} 5007	&	297.31 $\pm$  9.17	&	680.52 $\pm$ 20.45	&	397.17 $\pm$ 12.70	&	569.46 $\pm$ 17.26	&	272.65 $\pm$  8.75	& \\
{[}O~{\sc i}{]} 6300	&	  3.54 $\pm$  0.59	&	  1.82 $\pm$  0.10	&	  2.37 $\pm$  0.58	&	  2.39 $\pm$  0.23	&	  3.31 $\pm$  0.37	& \\
{[}S~{\sc iii}{]} 6312	&	  ---				&	  1.89 $\pm$  0.09	&	  0.00 $\pm$  0.00	&	  2.45 $\pm$  0.27	&	  1.18 $\pm$  0.31	& \\
{[}N~{\sc ii}{]} 6548	&	  ---				&	  1.74 $\pm$  0.09	&	  5.09 $\pm$  0.74	&	  1.90 $\pm$  0.20	&	  3.05 $\pm$  0.39	& \\
H$\alpha$ 6563	&	280.60 $\pm$  8.82	&	281.58 $\pm$  8.50	&	281.95 $\pm$  9.12	&	280.83 $\pm$  8.56	&	280.33 $\pm$  9.32	& \\
{[}N~{\sc ii}{]} 6584	&	  8.60 $\pm$  0.38	&	  5.32 $\pm$  0.20	&	 11.86 $\pm$  0.82	&	  6.42 $\pm$  0.35	&	  6.32 $\pm$  0.46	& \\
{[}S~{\sc ii}{]} 6716	&	 23.75 $\pm$  1.19	&	  9.01 $\pm$  0.31	&	 23.58 $\pm$  1.11	&	 14.28 $\pm$  0.58	&	 18.54 $\pm$  0.84	& \\
{[}S~{\sc ii}{]} 6731	&	 15.87 $\pm$  0.95	&	  6.84 $\pm$  0.24	&	 16.01 $\pm$  0.89	&	  9.80 $\pm$  0.45	&	 12.73 $\pm$  0.67	& \\
\midrule
H$\beta$ 4861		&	1.13E-14			&	7.17E-14			&	8.78E-15			&	2.10E-14			&	8.85E-15			& \\
EW(H$\alpha$)		&	211				&	811				&	1982				&	631				&	103				& \\
EW(H$\beta$)		&	47.9				&	177				&	710				&	106				&	30				& \\
c(H$\beta$)		&	0				&	0.071			&	0.180			&	0.127			&	0.115			& \\
\addlinespace[5pt]\bottomrule\addlinespace[5pt]}

\FloatBarrier 

\subsection{Notes on particular objects}
\begin{figure}[htbp]
\centering
\includegraphics[width=0.6\hsize]{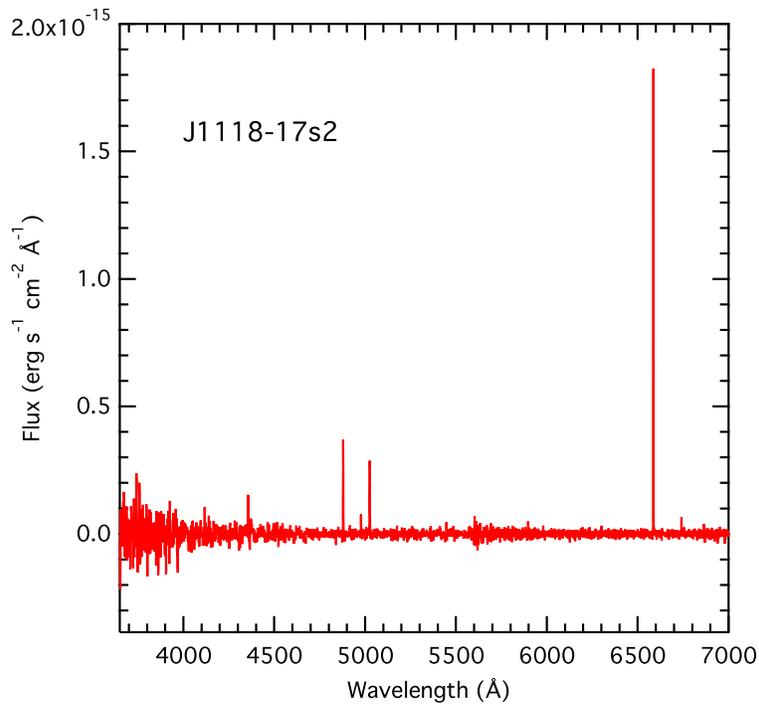}
\caption{Spectrum for J1118-17s2 (pipeline ``sag'' removed).}\label{fig_5}
\end{figure}

\begin{figure}[htbp]
\centering
\includegraphics[width=0.6\hsize]{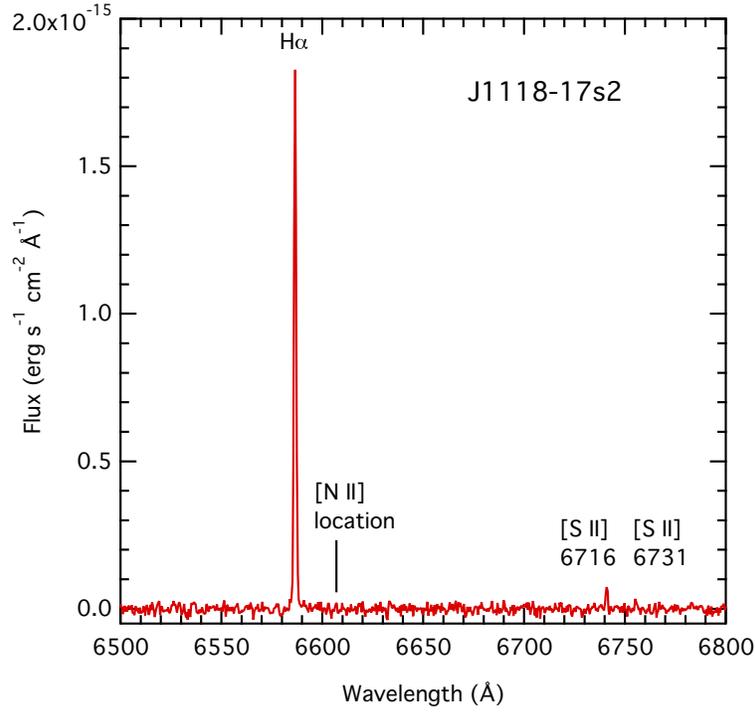}
\caption{Section of spectrum for J1118-17s2, showing apparent absence of any [N~{\sc ii}] emission lines.}\label{fig_6}
\end{figure}

\noindent The spectrum from \textbf{J1118-17s2} is shown in Figure \ref{fig_5}. An enlarged section of the spectrum of J1118-17s2 near H$\alpha$ is shown in Figure \ref{fig_6}, illustrating the apparent absence (to within the noise) of {[}N {\sc ii}{]}, although both {[}S {\sc ii}{]} lines are apparent. This lack of any evidence for nitrogen suggests a particularly low metallicity, which we have estimated below using strong line diagnostic measurements. This object (and the associated J1118-17s1) warrants further observation to reduce the noise and establish the {[}N {\sc ii}{]} flux.  We have not presented the spectrum of J1118-17s1 as the signal-to-noise ratio was very poor and did not permit reliable measurement of any fluxes other than H$\alpha$, H$\beta$ and [O~{\sc iii}] 5007 \AA.

\begin{figure}[htbp]
\centering
\includegraphics[width=0.9\hsize]{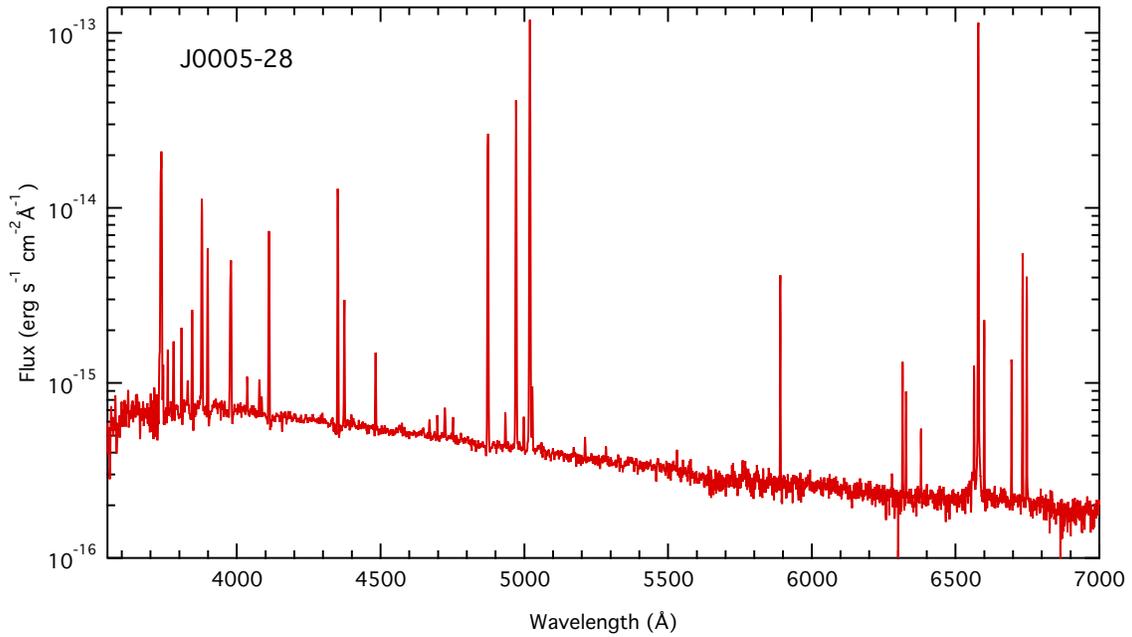}
\caption{Spectrum of J0005-28, log flux axis.  Several Balmer lines are apparent, and the [O~{\sc iii}] 4363\AA~ line is strong.}\label{fig_7}
\end{figure}

\noindent Figure \ref{fig_7} shows the emission line rich spectrum of the bright BCD \textbf{J0005-28} with flux on a logarithmic scale.  Twelve Balmer lines can be seen, allowing particularly accurate de-reddening to be calculated. The de-reddening process based on the H$\alpha$ to H$\beta$ ratio gave ratios to within 0.3\% for H$\delta$ and 12\% for H$\epsilon$ of the expected values for Case B.

\FloatBarrier

\section{Nebular metallicities}

\subsection{Electron temperatures and oxygen abundances}

The electron temperature, T$_e$, can be derived from collisionally excited line fluxes, for a variety of ionic species, provided the auroral line is observed (in the case of O~{\sc iii}, the 4363\AA~line). The method most frequently used makes use of the ratio of fluxes of the bright [O~{\sc iii}] lines to the auroral line.  This is a well-established technique \citep[see discussions in][]{2006agna.book.....O,2013ApJS..207...21N}, but it calculates only the O$^{++}$ abundance, not the total gas-phase abundance of oxygen.  In most \HII regions, the contributions to total oxygen from neutral O and O$^{+++}$ are minor, so in addition to O$^{++}$ we only need to calculate the contribution from O$^+$. 

If the equivalent auroral lines for [O~{\sc ii}], [N~{\sc ii}] and [S~{\sc iii}] are present in the spectra, the electron temperatures can be calculated using these lines too, and since they peak at different regions in the \HII region, the auroral lines collectively sample the complete volume. When these auroral lines are not observed, empirical methods can be used to estimate the O$^+$ abundance, for example in \citet[][Equations 3, 4]{2006A&A...448..955I}. However, as those authors note, the methods depend on having reliable atomic data (energy levels, transition probabilities, and collision strengths). Consequently we have approached the problem again from first principles, using the latest atomic data, to derive the total oxygen abundance \footnote{In this work, we do not have data for the [O~{\sc ii}] $\lambda\lambda$7320,30 lines, so the Izotov method provides a useful comparison.}.

The rate of collisional excitation for O$^{++}$ from the $^3$P ground state(s) to the $^1$D$_2$ level is given, for the thermal equilibrium case \citep{2012ApJ...752..148N}, by,

\begin{equation}\label{eq1}
 r_{12}=n_e n_{O^{++}}  \left( \frac{h^2  \sqrt{2}}{4 \pi^{3/2} m^{3/2}_e \sqrt{k_B}}\right)  \frac{1}{ g_1~T_e} \Upsilon_{12}(T) ~exp\left(-\frac{E_{12}}{k_BT_e}\right) ,
\end{equation}
where $h$ is the Planck constant, $m_e$ is the electron mass, $g_1$ is the statistical weight of the ground state (= 9 for O$^{++}$), $k_B$ is the Boltzmann constant, $\Upsilon_{12}$ is the net effective collision strength for collisional excitations from the ground $^3$P states to the $^1$D$_2$ state and E$_{12}$ is the energy level of the  $^1$D$_2$ state. Ignoring the small contribution to the population of the  $^1$D$_2$ level from radiative cascade from higher energy levels, the rate of emission of photons from that level is equal to the rate of excitation, i.e., $r_{12} = r_{21}$. The emissivity of [O~{\sc iii}] from transitions from the  $^1$D$_2$ ($\lambda\lambda$ 5007, 4959 and 4931) level is proportional to $r_{21} \times E_{12}$. Here we have used the total effective collision strengths for the forbidden $^1$D$_2$  to $^3$P transitions, so we use the flux-weighted photon energy, corresponding to a wavelength of 4997 \AA, for  $E_{12}$ .

The emissivity of H$\beta$ is proportional to $n_e \times n_{H^+} \times \alpha^{eff}_B (\text{H}_{\beta})$ \citep{2003adu..book.....D}, where $n_e$ is the electron density,  $n_{H^+}$ is the ionized hydrogen density, and $\alpha^{eff}_B (\text{H}_{\beta})$ is the effective emissivity for H$\beta$, which takes into account photon energies and branching ratios, and for which values have been computed by \cite{1995MNRAS.272...41S}.

Given that the ratio of the flux of [O~{\sc iii}] to that of H$\beta$ is equal to the ratio of their emissivities multiplied by their photon energies, for a given geometry, one may reorganise the above equations to derive an expression for the ratio of the number density of O$^{++}$ ions to hydrogen ions (i.e., the O$^{++}$ abundance) in terms of the flux ratio of [O~{\sc iii}]($^1$D$_2$) to H$\beta$,
\begin{equation}\label{eq2}
\frac{n_{O^{++}}}{n_{H^+}} = \frac{\text{flux}(\text{O$^{++}$)}}{\text{flux}(\text{H}_\beta)}.g_1.\sqrt{T_e}.\alpha^{eff}_B (\text{H}_{\beta}).\text{exp}(E_{12}/(kT_e)) \times115885.4/(E_{12}.\Upsilon_{12})
\end{equation}
where $T_e$ is the electron temperature derived from the [O~{\sc iii}] line ratio, for which there is a simple expression from \cite{2013ApJS..207...21N},
\begin{equation}\label{eq3}
T_e = a~(-log_{10}({\mathcal{R}}) -b)^{-c} ,
\end{equation}
where, for [O~{\sc iii}],
\begin{equation}\label{eq4}
\mathcal{R} = \frac{j(\lambda4363)}{j(\lambda5007)+j(\lambda4959)} , 
\end{equation}
and a= 13229, b= 0.92350, and c=0.98196.

In an identical fashion, one may derive an expression for the abundance of O~{\sc ii} using the observed fluxes from the [O~{\sc ii}] 3726,3729$\lambda\lambda$ lines,
\begin{equation}\label{eq5}
\frac{n_{O^+}}{n_{H^+}}  = \frac{\text{flux}(\text{O$^+$)}}{\text{flux}(\text{H}_\beta)}.g_{1(\text{O$^+$})}.\sqrt{T_e}.\alpha^{eff}_B (\text{H}_{\beta}).\text{exp}(E_{12(\text{O$^+$})}/(kT_e)) \times115885.4/(E_{12(\text{O$^+$})}.\Upsilon_{12(\text{O$^+$})}) ,
\end{equation}
where, in this case, $T_e$ is the electron temperature derived from the [O~{\sc ii}] ratio \citep[see][]{2013ApJS..207...21N} using the  ratio of the 7320,30$\lambda\lambda$ lines to the 3726,3729$\lambda\lambda$ lines.  If, as in the case of these observations, the NIR lines are not available, it is possible to derive an expression for the [O~{\sc ii}] electron temperature from the Mappings photoionization models as a polynomial in terms of \emph{total} gas-phase oxygen abundance,
\begin{equation}\label{eq6}
T_e([\text{O~{\sc ii}}])=T_e([\text{O~{\sc iii}}]) \times (3.0794 - 0.086924~Z - 0.1053~Z^2 + 0.010225~Z^3)
\end{equation}
where Z=12+log(O/H).

This does not provide the final answer, and it is necessary to iterate to a final value for the abundance of O$^+$, starting by using the O$^{++}$ abundance as the total oxygen abundance.  The process converges in less that five iterations. \cite{1992AJ....103.1330G} and  \cite{2012MNRAS.426.2630L} have used a simpler approach, expressing the low ionization zone temperature (effectively the [O~{\sc ii}] temperature) in terms of the [O~{\sc iii}] temperature, which does not require iteration. \cite{1992AJ....103.1330G} used a linear relation, whereas \cite{2012MNRAS.426.2630L} used a more complex fit to photoionization model data.

Equation \ref{eq6a} shows the expression used by \cite{2012MNRAS.426.2630L}:
\begin{equation}\label{eq6a}
T_e(\text{O{\sc ii}})=T_e(\text{O{\sc iii}}) +450 -70\times exp\left[(T_e(\text{O{\sc iii}})/5000)^{1.22}\right]
\end{equation}
Equation \ref{eq6a} gives total oxygen abundance values close to those from iterating Equation \ref{eq6}. Values determined for  oxygen abundances are not exact, because of the nature of the approximations used, the calculated values for oxygen abundances depend on the photoionization models used to build the models, and the use of a model derived from a single value of the ionization parameter, $q$. Testing the two methods (Equations \ref{eq6} and \ref{eq6a}) against artificial data indicates that they generate total oxygen abundances within 1\% of the input values. The iterative approach (Equation \ref{eq5}) is marginally the more consistent of the two over a range of ionization parameter values.

The above equations may be simplified for computation by using accurate expansions in terms of the [O~{\sc iii}] electron temperature to $\alpha^{eff}_B (\text{H}_{\beta})$, $\Upsilon (^1\text{D}_2)$, and $\Upsilon (^2\text{D}^0_{3/2,5/2})$.  The Case B emissivity data  for H$\beta$ as a function of temperature, from \cite{1995MNRAS.272...41S} may be fit with a simple power law,
\begin{equation}\label{eq7}
\alpha^{eff}_B (\text{H}_{\beta}) = -\text{1.7221e-26} + \text{1.4772e-22} \times \text{T}_e^{-0.75538}
\end{equation}

The effective collision strengths for the O$^{++}$ $^1$D$_2$ level \citep{2012MNRAS.423L..35P}, from which the $\lambda\lambda$4959, 5007 doublet originates, can be fit with a simple exponential function of temperature,
\begin{equation}\label{eq8}
\Upsilon (^1\text{D}_2) = 3.0733 - 0.94563 \times \text{exp}((5000-\text{T}_e)/12105) ,
\end{equation}
and the effective collision strengths for O$^{++}$ from \cite{2007ApJS..171..331T} for the composite upper state, $^2$D$^0_{3/2,5/2}$, from which the $\lambda\lambda$3726,9 doublet originates, can similarly be fit by a linear function of temperature,
\begin{equation}\label{eq9}
\Upsilon (^2\text{D}^0_{3/2,5/2}) = 1.3394 + \text{1.3443e-06} \times \text{T}_e.
\end{equation}

Applying these methods to the present observations, we obtain the electron temperatures and total gas-phase oxygen abundances shown in Table \ref{tx}. The values for J1118-16s2 are not listed as the $\lambda$4363 line was not observed. The uncertainties in T$_e$ were calculated from the flux error values, in Equation \ref{eq3}, and propagated through to the abundance values. See the appendix for a discussion of the error estimation.

\subsection{Strong-line diagnostic grids}

There are two principal methods for determining oxygen abundances from \HII region optical spectra, the direct or electron temperature (T$_e$) method and the strong line methods. The T$_e$ method is possible if one of the auroral lines is observed at adequate signal-to-noise ($> 3\sigma$), usually [O III] 4363\AA.  This is the case in all but one of the galaxies discussed here and is detailed below.  The so-called strong line methods use flux ratios of the prominent nebular lines to determine abundances \citep[e.g.,][]{2013ApJS..208...10D, 2002ApJS..142...35K, 2008ApJ...681.1183K}. Conventionally, the strong line methods were empirical, calibrated against results using the direct method. However, recently, \cite{2013ApJS..208...10D} have extensively revised the strong line techniques, developing diagnostic grids based on the Mappings photoionization modelling code, using the latest atomic data, and the possibility that the electrons exhibit a non-equilibrium $\kappa$ energy distribution \citep{2012ApJ...752..148N,2013ApJS..207...21N}. The grids  are new, and the ratios used have been selected to maximise the orthogonality of the parameters, avoiding to a large extent the degeneracy of older diagnostics, and to solve for both metallicity and ionization parameter. These new diagnostics generate values for both the oxygen abundance and the ionization parameter, $q$, and give substantially more consistent abundance values than the older methods. This can be seen by comparing the metallicity results for the different diagnostics from Table \ref{t3} and the older diagnostic results listed in Table \ref{tx}. In Table \ref{t3}, the diagnostics involving the ratio [N~{\sc ii}]/[S~{\sc ii}] and  [N~{\sc ii}]/[O~{\sc ii}] are particularly consistent, differing by typically $<$0.03 dex.

The ionization parameter $q$ (sometimes expressed as $U = q/c$, where $c$ is the speed of light) is the ratio of the number of ionizing source photons passing through a unit volume to the neutral hydrogen density.  The photon flux matches the number of new ions it produces, and as $q$ has the dimensions of velocity, it can be understood as the maximum speed at which the boundary of the ionized region can move outwards \citep{2003adu..book.....D}. $q$ is at its maximum at the inner edge of the ionized region of an \HII region, and falls to zero at the outer edge of the ionized nebula, where the ionizing flux is fully depleted. A problem with older diagnostics such as R$_{23}$ is that measured metallicities depend on the ionization parameter. There have been previous attempts to solve for the ionization parameter \citep{1994ApJ...426..135M,2002ApJS..142...35K,2005ApJ...631..231P} but  these new diagnostics solve for its value independently of the metallicity, and consequently, take into account the ionization gradients present in \HII regions. The diagnostics chosen here are also relatively insensitive to non-equilibrium ($\kappa$) electron energy distributions, especially for values of $\kappa > $ 20. In addition, $\kappa$ distributions have a smaller effect on the excitation of lower metallicity \HII regions than in higher metallicity objects. For this reason, and in the interests of clarity, we present here only the equilibrium (Maxwell-Boltzmann, or $\kappa=\infty$) results.

\begin{figure}[htbp]
\centering
\includegraphics[width=0.9\hsize]{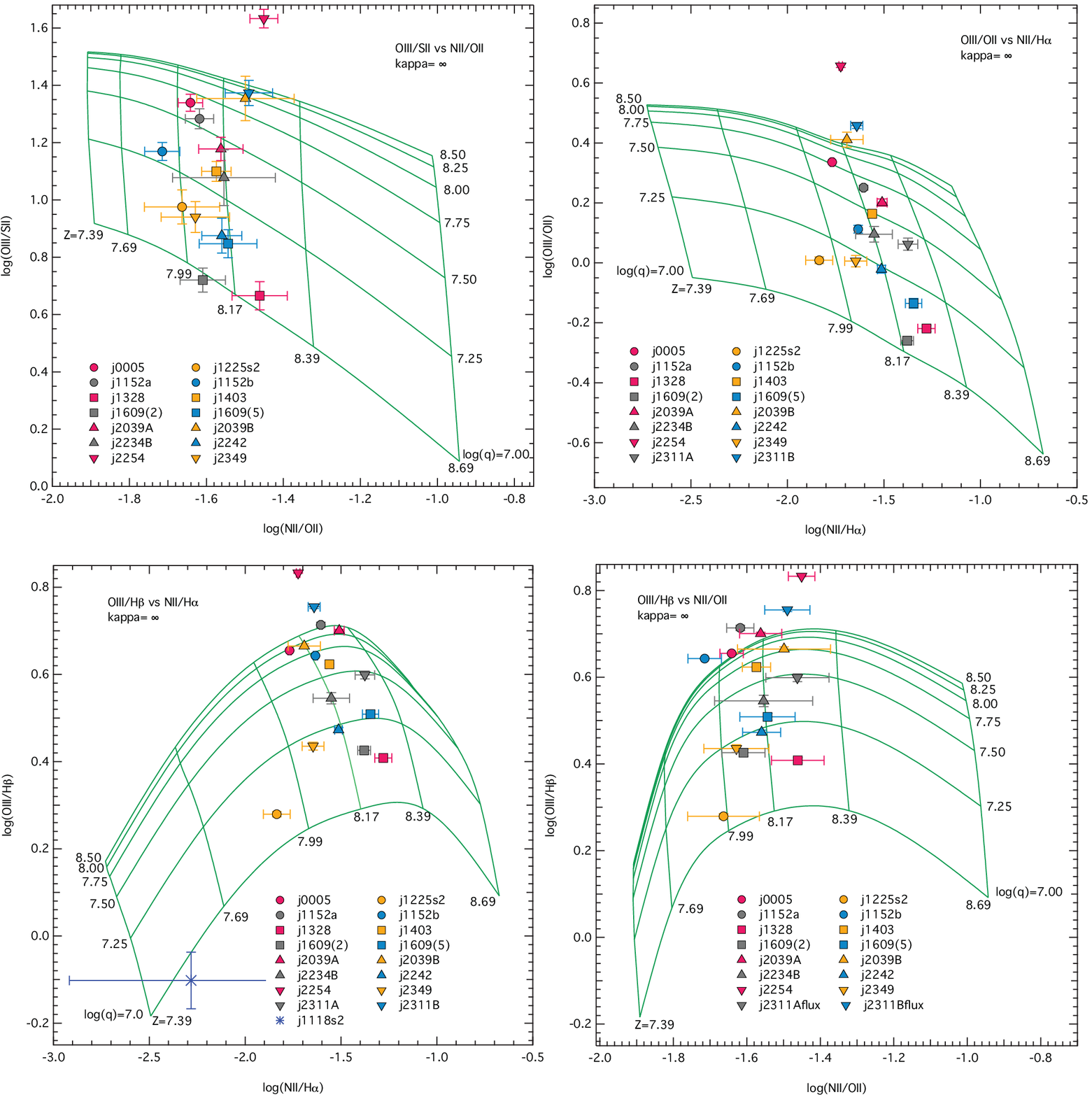}
\caption{Observed flux ratios for SIGRID objects plotted on the OIII/SII--NII/OII, OIII/OII--NII/Ha, OIII/Hb--NII/Ha and OIII/Hb--NII/OII grids }\label{fig_8}
\end{figure}

\begin{figure}[htbp]
\centering
\includegraphics[width=0.9\hsize]{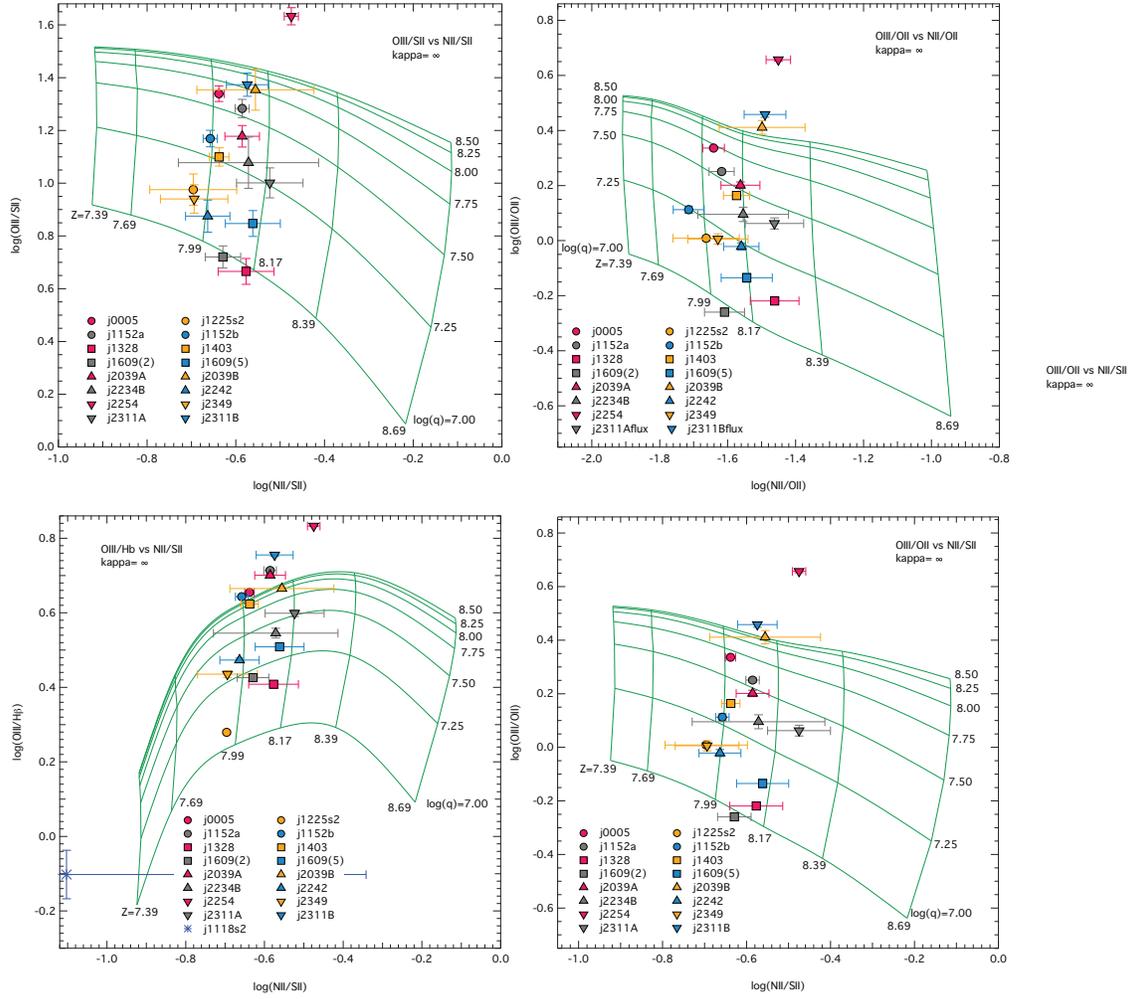}
\caption{Observed flux ratios for SIGRID objects plotted on the OIII/SII--NII/SII, OIII/OII--NII/OII, OIII/Hb--NII/SII and OIII/OII--NII/SII grids }\label{fig_9}
\end{figure}

Figures \ref{fig_8} and \ref{fig_9} plot the log flux ratios for the observed objects on the diagnostic grids from \cite{2013ApJS..208...10D}. Not all of the objects can be accommodated within these grids.  There are several possible reasons, and these will be discussed in detail in the second paper in this series. One likely cause relates to electron densities. The grids shown here are calculated for an electron density, n$_e$ $\sim$ 5 cm$^{-3}$.  For the majority of the observed objects, this is accurate, but for some  the densities are somewhat higher.  Below we analyse the ratios of the two [S~{\sc ii}] lines (6716 and 6731 \AA) which are a useful diagnostic of electron density \citep{2006agna.book.....O}, and it is clear that some of the objects exhibit higher densities.  Some of the ``misfit'' points can be accommodated on grids calculated for higher electron densities (see below and Figure \ref{fig_11}).

\FloatBarrier 
\subsection{Strong Line metallicities}

Table \ref{t3} lists the oxygen abundances and ionization coefficients computed from the new diagnostic grids using the ``pyqz'' interpolation described in \cite{2013ApJS..208...10D}. While the interpolation scheme does not always work reliably for near-vertical grid lines, leading to null results, it is clear that different diagnostics yield somewhat different results.  However, the consistency is far better than earlier methods permitted.  We have found that for low metallicity objects ($<$ 0.5 Z$_\odot$), diagnostics listed in Table \ref{t3} involving log(N~{\sc ii}/S~{\sc ii}) give values for the metallicity that differ by typically less than 0.02 dex and diagnostic using the log(N~{\sc ii}/O~{\sc ii}) ratios are similar. It is also evident from Table  \ref{t3} that the direct methods are nearly all lower by $\sim$0.13 dex than the best log(N~{\sc ii}/S~{\sc ii}) strong line diagnostic values.  This is in agreement with the findings of \cite{2012MNRAS.426.2630L} that the direct method abundances are generally lower than strong line estimates.  However, with the newer atomic data, the recalculated direct method abundances, and the revised Mappings photoionization code, these differences are smaller. For comparison we also present the results of older strong line diagnostics, in Table \ref{tx}.  Perhaps the most variable result is that for J1118-17s2.  This is not surprising, as the [N~{\sc ii}] flux is poorly defined. It appears likely that an oxygen abundance figure of $\sim$7.2 ($\sim$Z$_\odot$/30) is a reasonable estimate.

It is worth noting that the Mappings photoionization modelling grids used here take into account the total oxygen abundance, i.e., both the gas-phase oxygen and that incorporated in dust grains. When comparing the electron temperature and strong line  abundances, it is necessary to increase the electron temperature oxygen abundance values by $\sim$0.07dex, to allow for the oxygen in dust grains that the direct method does not account for. This, of course, assumes a particular level of dust in the ISM.  In the Mappings strong line diagnostic grids, we assume a 1.0 solar dust depletion. \cite{2014A&A...563A..31R} found that the gas-to-dust ratio varies considerably between objects, depending on their individual star formation histories, particularly at the low end of the galaxy mass scale. This is a further complication for any attempt to estimate the total oxygen abundance in an \HII region.  \cite{2010A&A...521A..63L} has shown a correlation between the reddening coefficient c(H$\beta$) and the gas-to-dust ratio in Wolf-Rayet galaxies, and it is likely that a similar relation holds for smaller dwarf galaxies. (We do not have the necessary FIR data to allow us to calculate the dust mass here). In the meantime, the direct method oxygen abundance measurements provide a lower limit to the total oxygen.

\ctable[
caption= {Metallicity results from Strong Line grids for $\kappa=\infty$. The diagnostic grids are described in detail in \cite{2013ApJS..208...10D}. The uncertainties quoted are based on the variance of the average  values for the first five diagnostics listed here.},
pos=h,
captionskip=3pt,
width=\textwidth,
label=t3,
doinside=\scriptsize
]{lllllllll}
{
\tnote[]{``--'' indicates the diagnostic does not return a value for abundance or ionization parameter.}
} 
{\toprule[2pt]
Diagnostic: & NII/SII vs &  NII/SII vs &  NII/OII vs &  NII/OII vs &  NII/SII vs &  NII/OII vs &  NII/Ha vs &  NII/Ha vs \\
 & OIII/SII & OIII/Hb & OIII/OII & OIII/SII & OIII/OII & OIII/Hb & OIII/Hb & OIII/OII \\
\midrule
J0005-28 & & & & & & & & \\
z  & 8.012 & -- & 8.0428 & 8.0355 & 8.0114 & -- & 8.1191 & 8.1011 \\
 $\log(q)$  & 7.6328 & -- & 7.7212 & 7.6596 & 7.6845 & -- & 7.9568 & 7.7802 \\
mean z & 8.012$\pm$0.000 & & & & & & & \\
mean  $\log(q)$ & 7.659$\pm$0.037 & & & & & & & \\
\midrule
J1118-17s2 & & & & & & & & \\
Z & -- & -- & -- & -- & -- & -- & 7.5448 & 7.4741 \\
 $\log(q)$ & -- & -- & -- & -- & -- & -- & 6.9500 & 6.7092 \\
mean z  & 7.509$\pm$0.050 & & & & & &  \\
mean  $\log(q)$  & 6.830$\pm$0.170 & & & & & &  \\%
\midrule
J1152-02A & & & & & & & & \\
z  & 8.0896 & -- & 8.0762 & 8.0736 & 8.0868 & -- & -- & 8.198 \\
$\log(q)$  & 7.5707 & -- & 7.5506 & 7.5611 & 7.558 & -- & -- & 7.6697 \\
mean z & 8.088$\pm$0.002 & & & & & & & \\
mean $\log(q)$ & 7.564$\pm$0.009 & & & & & & & \\
\midrule
J1152-02B & & & & & & & & \\
z  & 7.9802 & -- & 7.8965 & 7.9038 & 7.9819 & -- & 8.1878 & 8.118 \\
$\log(q)$  & 7.3404 & -- & 7.2459 & 7.3072 & 7.2869 & -- & 7.7019 & 7.3713 \\
mean z & 7.981$\pm$0.001 & & & & & & & \\
mean $\log(q)$ & 7.314$\pm$0.038 & & & & & & & \\
\midrule
J1225-06s2 & & & & & & & & \\
z  & 7.9289 & 7.9321 & 7.9924 & 7.9891 & 7.9256 & 7.9736 & 7.9106 & 7.9261 \\
 $\log(q)$  & 7.1382 & 7.0849 & 7.1958 & 7.157 & 7.1611 & 7.0573 & 7.0925 & 7.1651 \\
mean z & 7.929$\pm$0.003 & & & & & & & \\
mean  $\log(q)$ & 7.128$\pm$0.039 & & & & & & & \\
\midrule
J1328+02 & & & & & & & & \\
z  & 8.1465 & 8.1159 & 8.2471 & 8.2413 & 8.1337 & 8.2524 & 8.3003 & 8.2909 \\
$\log(q)$  & 6.9894 & 7.1697 & 7.0976 & 7.0408 & 7.0481 & 7.1428 & 7.0164 & 7.123 \\
mean z & 8.132$\pm$0.015 & & & & & & & \\
mean $\log(q)$ & 7.069$\pm$0.092 & & & & & & & \\
\midrule
J1403-27 & & & & & & & & \\
z  & 8.0137 & 8.0161 & 8.1393 & 8.1363 & 8.0106 & -- & 8.2286 & 8.1978 \\
 $\log(q)$  & 7.2684 & 8.045 & 7.4556 & 7.343 & 7.3739 & -- & 7.5551 & 7.4934 \\
mean z & 8.013$\pm$0.003 & & & & & & & \\
mean  $\log(q)$ & 7.562$\pm$0.421 & & & & & & & \\
 \midrule
J1609-04(2) & & & & & & & & \\
z  & 8.0654 & 8.0303 & 8.0418 & 8.0447 & 8.0681 & 8.0722 & 8.243 & 8.1959 \\
 $\log(q)$  & 6.992 & 7.2226 & 6.9825 & 6.9856 & 6.9843 & 7.2026 & -- & 7.0429 \\
mean z & 8.055$\pm$0.021 & & & & & & & \\
mean  $\log(q)$ & 7.066$\pm$0.135 & & & & & & & \\
\addlinespace[5pt]\bottomrule[1.5pt]\addlinespace[5pt]}

\ctable[
caption= {Metallicity results from Strong Line grids for $\kappa=\infty$},
pos=h,
captionskip=3pt,
width=\textwidth,
label=t3b,
doinside=\scriptsize,
continued
]{lllllllll}
{
\tnote[]{``--'' indicates the diagnostic does not return a value for abundance or ionization parameter.}
} 
{\toprule[2pt]
Diagnostic: & NII/SII vs &  NII/SII vs &  NII/OII vs &  NII/OII vs &  NII/SII vs &  NII/OII vs &  NII/Ha vs &  NII/Ha vs \\
 & OIII/SII & OIII/Hb & OIII/OII & OIII/SII & OIII/OII & OIII/Hb & OIII/Hb & OIII/OII \\
 \midrule
J1609-04(5) & & & & & & & & \\
z  & 8.1435 & 8.1261 & 8.1628 & 8.1616 & 8.1424 & 8.1753 & 8.3091 & 8.2626 \\
 $\log(q)$  & 7.1225 & 7.3364 & 7.1189 & 7.1261 & 7.129 & 7.3023 & 7.1813 & 7.182 \\
mean z & 8.137$\pm$0.010 & & & & & & & \\
mean  $\log(q)$ & 7.196$\pm$0.122 & & & & & & & \\
\midrule
J2039-63A & & & & & & & & \\
z  & 8.0889 & -- & 8.1589 & 8.1578 & 8.0853 & -- & 8.35 & 8.2661 \\
 $\log(q)$  & 7.4083 & -- & 7.5284 & 7.4515 & 7.4722 & -- & 8.3704 & 7.6159 \\
mean z & 8.087$\pm$0.003 & & & & & & & \\
mean  $\log(q)$ & 7.440$\pm$0.045 & & & & & & & \\
\midrule
J2039-63B & & & & & & & & \\
z  & 8.1379 & 8.1515 & -- & 8.2147 & -- & -- & 8.1677 & -- \\
 $\log(q)$  & 7.7838 & 8.127 & -- & 7.9763 & -- & -- & 7.9284 & -- \\
mean z & 8.145$\pm$0.010 & & & & & & & \\
mean  $\log(q)$ & 7.955$\pm$0.243 & & & & & & & \\
\midrule
J2234-04B & & & & & & & & \\
z  & 8.1115 & 8.1104 & 8.1667 & 8.1643 & 8.1079 & 8.1659 & 8.1713 & 8.1742 \\
 $\log(q)$  & 7.3037 & 7.4352 & 7.368 & 7.3359 & 7.3459 & 7.3929 & 7.0618 & 7.3863 \\
mean z & 8.110$\pm$0.002 & & & & & & & \\
mean  $\log(q)$ & 7.362$\pm$0.067 & & & & & & & \\
\midrule
J2242-06 & & & & & & & & \\
z  & 7.9952 & 7.9712 & 8.1484 & 8.1391 & 7.9836 & 8.15 & 8.16 & 8.1612 \\
 $\log(q)$  & 7.0778 & 7.3706 & 7.2361 & 7.1386 & 7.1563 & 7.2384 & 6.7332 & 7.2408 \\
mean z & 7.983$\pm$0.012 & & & & & & & \\
mean  $\log(q)$ & 7.202$\pm$0.152 & & & & & & & \\
\midrule
J2254-26 & & & & & & & & \\
no results & & & & & & & & \\
\midrule
J2311-42A & & & & & & & & \\
z  & 8.1817 & 8.1705 & 8.2661 & 8.2633 & 8.1760 & -- & 8.3488 & 8.3167 \\
 $\log(q)$  & 7.2614 & 7.5276 & 7.3981 & 7.3311 & 7.3515 & -- & 7.5032 & 7.4314 \\
mean z & 8.176$\pm$0.006 & & & & & & & \\
mean  $\log(q)$ & 7.380$\pm$0.135 & & & & & & & \\
\midrule
J2311-42B & & & & & & & & \\
z  & 8.1222 & -- & -- & 8.2207 & -- & -- & -- & -- \\
 $\log(q)$  & 7.8286 & -- & -- & 8.1406 & -- & -- & -- & -- \\
mean z & 8.171$\pm$0.049 & & & & & & & \\
mean  $\log(q)$ & 7.985$\pm$0.156 & & & & & & & \\
\midrule
J2349-22 & & & & & & & & \\
z  & 7.9360 & 7.9091 & 8.0443 & 8.0398 & 7.9293 & 8.0464 & 8.0732 & 8.0699 \\
 $\log(q)$  & 7.1122 & 7.3623 & 7.2168 & 7.1481 & 7.1602 & 7.2239 & 6.656 & 7.2224 \\
mean z & 7.925$\pm$0.014 & & & & & & & \\
mean  $\log(q)$ & 7.212$\pm$0.133 & & & & & & & \\
\addlinespace[5pt]\bottomrule[1.5pt]\addlinespace[5pt]}
\FloatBarrier 

\ctable[
caption= {{[O~{\sc iii}]} electron temperatures and gas-phase oxygen abundances},
doinside=\small,
pos=h,
sideways,
captionskip=3pt,
label=tx,
]{llcccccccc}
{
\tnote[1]{Z(this work) derived from Equations \ref{eq2}, \ref{eq5}, and \ref{eq6}}
\tnote[2]{The Z(grids) values are the average of the new grids involving the log(N~{\sc ii}/S~{\sc ii}) and log(N~{\sc ii}/O~{\sc ii}) ratios.}
\tnote[3]{$\delta$Z is the difference between Z(grids) and Z(this work).}
\tnote[4]{Older strong line methods (columns 7-10) described in \cite{2008ApJ...681.1183K}}
}
{\toprule[1.5pt]
Object		&	T$_e$ (K)			&	Z\tmark[ 1]		&	Z		& Z(grids)\tmark[2] &	$\delta$Z\tmark[3]	&	\multicolumn{4}{c}{Old strong line\tmark[4]}\\
			&					&	(this work)			&	(Izotov06)	&		&				&	 M91 	&  KK04	& PP04	& PP04	\\
			&					&					&			&		&				&			&		& 5007.N2 & N2	\\	
J0005-28 		& 14720 $\pm$ 36 		& 7.847 $\pm$ 0.025		& 7.858 		& 8.012 	&	 0.165 		& 8.104 		& 8.306 	& 7.954 	& 7.951 \\
J1152-02A 	& 12249 $\pm$ 34		& 8.151 $\pm$ 0.025		& 8.178 		& 8.088 	&	-0.063 		& 8.297 		& 8.466 	& 7.988 	& 8.034 \\
J1152-02B 	& 12723 $\pm$ 55		& 8.094 $\pm$ 0.026		& 8.108 		& 7.981 	&	-0.113 		& 8.288 		& 8.455	& 8.001 	& 8.021 \\
J1225-06s2 	& 16560 $\pm$ 395		& 7.498 $\pm$ 0.028		& 7.462 		& 7.929 	&	 0.431 		& 7.750 		& 7.999 	& 8.053 	& 7.910 \\
J1328+02 	& 14847 $\pm$ 464		& 7.867 $\pm$ 0.027		& 7.835 		& 8.132 	&	 0.265 		& 8.234 		& 8.397 	& 8.190 	& 8.165 \\
J1403-27 		& 14022 $\pm$ 71		& 7.939 $\pm$ 0.025		& 7.942 		& 8.013 	&	 0.074 		& 8.199 		& 8.381 	& 8.031 	& 8.054 \\
J1609-04b2 	& 10432 $\pm$ 491		& 8.345 $\pm$ 0.026		& 8.393 		& 8.055 	&	-0.290 		& 8.323 		& 8.471 	& 8.152 	& 8.127 \\
J1609-04b5 	& 14235 $\pm$ 984 		& 7.959 $\pm$ 0.028		& 7.944 		& 8.137 	&	 0.178 		& 8.313 		& 8.468 	& 8.136 	& 8.140 \\
J2039-63A 	& 14384 $\pm$ 182		& 7.965 $\pm$ 0.027		& 7.976 		& 8.087 	&	 0.121 		& 8.321 		& 8.485 	& 8.023 	& 8.076 \\
J2039-63B 	& 13984 $\pm$ 499		& 7.894 $\pm$ 0.029		& 7.911 		& 8.145 	&	 0.251 		& 8.066 		& 8.276 	& 7.976 	& 7.992 \\
J2234-04B 	& 14024 $\pm$ 754 		& 7.897 $\pm$ 0.032		& 7.890 		& 8.110 	&	 0.213 		& 8.118 		& 8.312 	& 8.059 	& 8.058 \\
J2242-06 		& 13625 $\pm$ 575		& 7.921 $\pm$ 0.027		& 7.906 		& 7.983 	&	 0.062 		& 8.109 		& 8.300 	& 8.094 	& 8.074 \\
J2254-26 		& 12871 $\pm$ 43		& 8.090 $\pm$ 0.025		& 8.129 		& --- 		& 	 --- 			& 8.250 		& 8.433 	& 7.912 	& 7.976 \\
J2311-42A 	& 12584 $\pm$ 580		& 8.090 $\pm$ 0.029		& 8.100 		& 8.176 	&	 0.086 		& 8.254 		& 8.425 	& 8.098 	& 8.129 \\
J2311-42B 	& 13453 $\pm$ 160		& 8.011 $\pm$ 0.027		& 8.038 		& 8.171 	&	 0.160 		& 8.217 		& 8.403 	& 7.963 	& 8.018 \\
J2349-22 		& 13837 $\pm$ 1054 	& 7.858 $\pm$ 0.029		& 7.839 		& 7.925 	&	 0.067 		& 8.015 		& 8.222 	& 8.064 	& 8.015 \\
\addlinespace[5pt]\bottomrule[1.5pt]\addlinespace[5pt]}

\FloatBarrier

\subsection{log(N/O)}

In this section we use the approach from our previous paper, \cite{2014ApJ...780...88N}. One of the more important parameters in understanding galactic evolution is the nitrogen abundance, and in particular, the ratio of nitrogen to oxygen.  The observations reported here include relatively low noise measurements of both [N~{\sc ii}] and [O~{\sc ii}], allowing us to explore the values of log(N/O) for each \HII region.  To calculate the value of N/O from [N~{\sc ii}] and [O~{\sc ii}] line fluxes, we use empirical formulae from \cite{2006A&A...448..955I}, Equations (3) and (6).  (This approach was chosen because it accounts for the temperature dependencies of the [N~{\sc ii}] and [O~{\sc ii}] fluxes). These equations reduce to:

\begin{equation}\label{eq10}
\textrm{ log }\left(\frac{N}{O}\right) = \textrm{ log }\left(\frac{NII ~6584+6548}{OII ~3726+3729} \right) + 0.273 - 0.726/T_{e4} +0.007*T_{e4} - 0.02*\textrm{ log }(T_{e4}) ,
\end{equation}
where $T_{e4}$ is the [O~{\sc iii}] electron temperature in units of 10,000K. This equation differs only by a small constant offset (0.033) from that quoted by \citet[Equation (9)]{1992MNRAS.255..325P}, most probably due to the latter using older atomic data.. We assume the same electron temperature for  O~{\sc ii} and N~{\sc ii} (reasonable, as they both arise primarily from the outer parts of the \HII region), and further, that N$^+$/O$^+$ = N/O, following \cite{2010ApJ...720.1738P} and others. The errors from these assumptions are likely to be of the same order as the measurement uncertainties. The results are shown in Table \ref{ty}.

\begin{figure}[htbp]
\centering
\includegraphics[width=0.8\hsize]{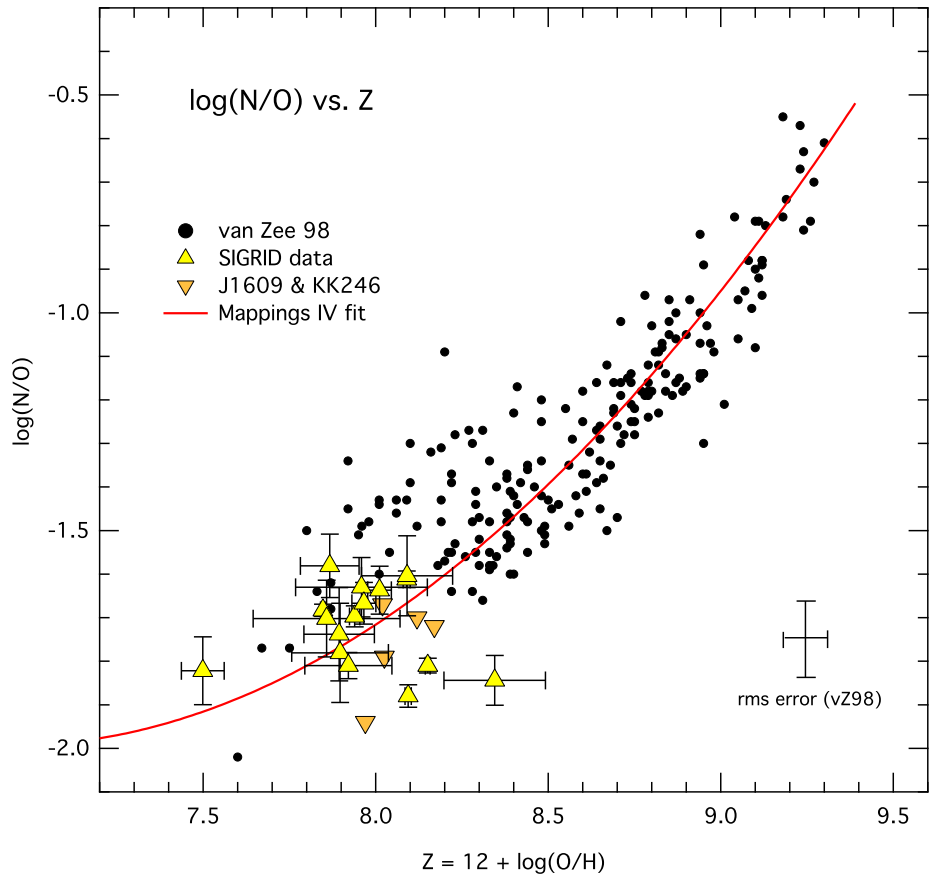}
\caption{Log(N/O) versus oxygen abundance, Z.  The triangles are for SIGRID objects, the black points are from \citet{1998ApJ...497L...1V}. The lower metallicity van Zee data are shown only for dwarf galaxies for which electron temperature metallicities are available.}\label{fig_10}
\end{figure}

Using the abundance values listed in Table \ref{ty}, we can plot log(N/O) versus oxygen abundance. Figure \ref{fig_10} shows the data from this work  (yellow triangles), data from other SIGRID objects from \cite{2014ApJ...780...88N} (brown triangles) and data from \cite{1998AJ....116.2805V} (black circles). The SIGRID data are consistent with the van Zee results, without any obvious evidence of a floor.  However, \cite{2012ApJ...754...98B} state that the nitrogen floor does not become apparent until the oxygen abundance falls below Z=7.7, so the SIGRID data do not resolve the question of whether the floor exists.  While the data for J1118-17s2 are not plotted because the value of the [N~{\sc ii}] flux is not well defined, the best estimate values for this object (log(N/O) $\lesssim$ 2.481 and Z $\lesssim$ 7.2) extend the trend considerably further in the same direction, off the graph, below and to the left. If correct, this suggests a very low level for any primary nitrogen, but further observations are necessary to confirm this The red curve is the fit to the van Zee data used in the Mappings model grids, from \citet[][Figure 3]{2013ApJS..208...10D}, making allowance for the oxygen depletion into dust grains. 

There is an increased scatter in the distribution with decreasing oxygen abundance (Z).  Two possibilities may contribute to this.  First, as there are populations of older stars in these galaxies \citep[for example, KK246,][]{2014ApJ...780...88N}, so intermediate mass AGB stars will contribute nitrogen to the interstellar medium (ISM) through hot-bottom burning processes. Second, the amounts contributed by such processes will depend on the (unknown) star formation histories of different galaxies. As the abundances derived using the strong line diagnostics depend at least in part on the log(N~{\sc ii}/O~{\sc ii}) and log(N~{\sc ii}/S~{\sc ii}) ratios, the results are sensitive to deviations in the nitrogen fit from the theoretical fit used in the Mappings models, which was derived from van Zee's 1998 data \citep[see][]{2013ApJS..208...10D}, so any error here affects the model outcomes.

Figure \ref{fig_10} shows that, at lower metallicities,  the data exhibit increasing scatter and may even have started to fork into two branches. The upper region of the scatter may indicate nitrogen enrichment by Wolf-Rayet  WN stars, as suggested by \cite{2010A&A...517A..85L}.  Smaller galaxies may divide into two classes, those with (or that have had) WN stars, and those without, depending on the stochastic nature of individual star formation events. \cite{1978MNRAS.185P..77E} have suggested the [N/O] ratio of \HII regions in a galaxy arise from nitrogen that is significantly primary in origin, and are a measure of the early star formation history.  While this may be correct for larger galaxies, at least in the case of the very isolated dwarf galaxy KK246, it is not the case, as the log(N/O) ratio is low but there is evidence of older stellar populations \citep{2014ApJ...780...88N}. The presence or absence of WN stars in a dwarf galaxy's \HII regions is a plausible explanation for this scatter or bifurcation.  This would be consistent with the observations of the Blue Compact Dwarf galaxy, HS0837+4717 \citep{2004A&A...419..469P, 2011A&A...532A.141P}. The object appears to harbor over 100 Wolf-Rayet stars and has both a very low oxygen abundance and a high nitrogen abundance.

\FloatBarrier 
\subsection{SII line ratios}

The flux ratios of the two [S {\sc ii}] lines at 6716 and 6731 \AA~ are good indicators of electron density \citep[section 5.6,][]{2006agna.book.....O}.  Table \ref{t8} shows the variation of this ratio calculated for electron densities n$_e$ of  $\sim$ 5 and $\sim$50 cm$^{-3}$, for an ionization parameter  $\log(q)$ = 7.5, typical for \HII regions.  The trends in the [S~{\sc ii} 6716]/[S~{\sc ii} 6731] line ratio are due to two factors: (1) the relatively small dependence of the line ratio on n$_e  \sqrt{\textrm{T}_e}$ resulting from the collisional excitation rates of the S~{\sc ii} line upper states, and (2) the use of the isobaric setting in the Mappings photoionization modelling code, such that the density structure of the S~{\sc ii} region is a function of the (varying) temperature within it, which depends on metallicity.

\ctable[
caption= {Calculated {[}S {\sc ii}{]} line ratios vs gas-phase oxygen abundance (Z=log(O/H)) for n$_e$= 10 and 100 cm$^{-3}$ and ionization parameter  $\log(q)$=7.5},
doinside=\small,
pos=h,
captionskip=3pt,
label=t8,
]{lcccccc}
{\tnote[]{Calculated using Mappings IV photoionization code \citep{2013ApJS..208...10D}}.}
{\toprule[1.5pt]
Z			&	7.39		&	7.69		&	7.99		&	8.17		&	8.39		&	8.69 \\
\midrule
n$_e$=10		&	1.439	&	1.439	&	1.439	&	1.440	&	1.441	&	1.445 \\
\midrule
n$_e$=100	&	1.394	&	1.393	&	1.391	&	1.389	&	1.384	&	1.373 \\
 \addlinespace[5pt]\bottomrule[1.5pt]\addlinespace[5pt]}

Table \ref{ty} shows the measured [S {\sc ii}] line ratios and electron densities, calculated using PyNeb \citep{2012IAUS..283..422L}, for all objects except J1118-17s2, for which we have no electron temperature.  Comparing the observed S~{\sc ii}  line flux ratios to Table \ref{t8}, it is reasonable to conclude that the Mappings values show J0005-28, J1152-02A\&B, J1403-27, J2039-63A, J2234-04B, J2254-26 have electron densities n$_e>$ 5 cm$^{-3}$, while the remainder have n$_e<$ 5 cm$^{-3}$. This is confirmed quantitatively using PyNeb to estimate the actual electron densities.

\begin{figure}[htbp]
\centering
\includegraphics[width=0.8\hsize]{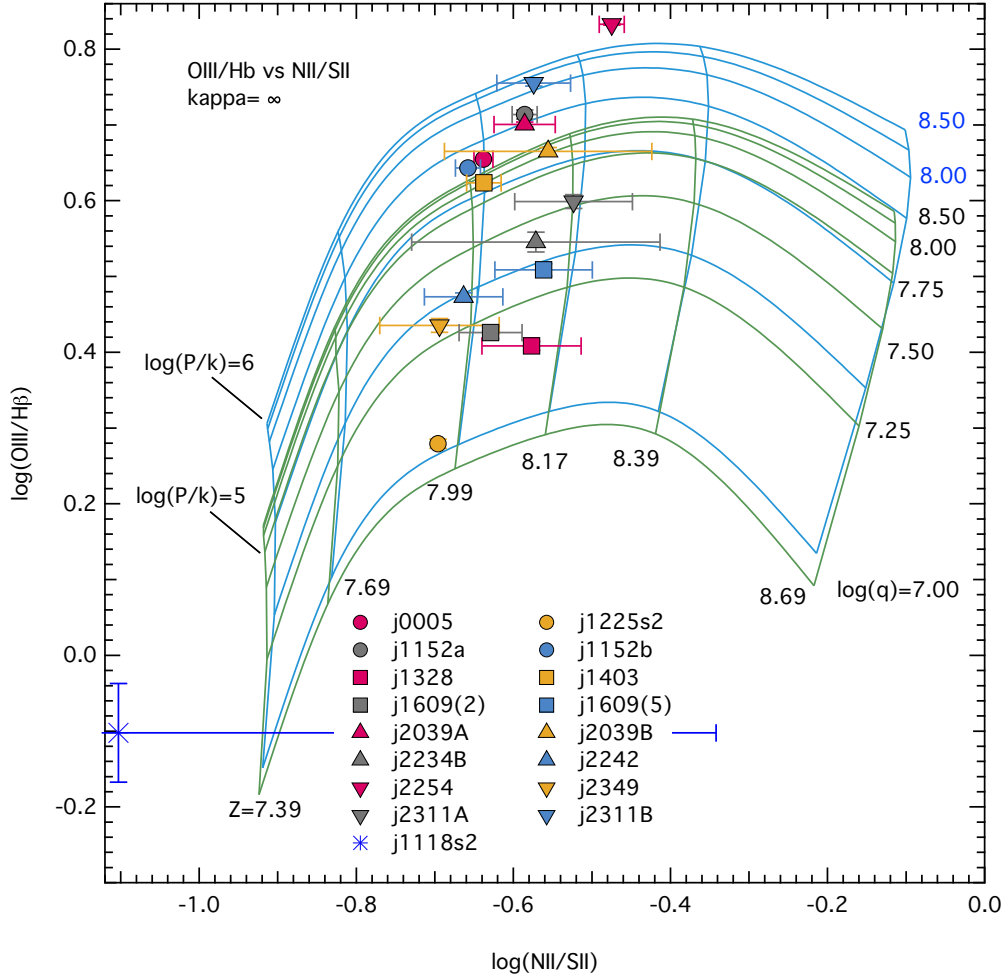}
\caption{Comparison of diagnostic grids for the ratios log(OIII/H$\beta$) versus log(NII/OII) for electron densities n$_e\sim$5 and $\sim$50 cm$^{-3}$ (isobaric case,  $\log(P/k)$ = 5 and 6, respectively, where $P$ is the pressure and $k$ is the Boltzmann constant.)}\label{fig_11}
\end{figure}

Figure \ref{fig_11} shows the diagnostic grids for O~{\sc iii}/H$\beta$ versus N~{\sc ii}/S~{\sc ii} at the two electron densities---the blue (upper) grid is for n$_e\sim$ 50 cm$^{-3}$, the green (lower) is for $\sim$5 cm$^{-3}$.  It is clear that all but J2254-26 can be accommodated even on the higher electron density grid.  The abundances for each object are very similar on both grids, but the estimated ionization parameter  $\log(q)$ changes. Similar results apply for the other diagnostic grids.  It is interesting to note that J2254-26 has the highest calculated electron density of the observed objects.

\ctable[
caption= {Log(N/O), log(N~{\sc ii}/O~{\sc ii}), {[S~{\sc ii}]} line ratios and electron densities.},
doinside=\small,
pos=h,
sideways,
captionskip=3pt,
label=ty,
]{lllll}
{
\tnote[1]{log(N/O) calculated from N{\sc ii}/O{\scshape ii} flux ratios using Equation \ref{eq10} and electron temperatures from Table \ref{tx}}.
\tnote[2]{n$_e$ uncertainties calculated using the line ratio uncertainties, except where these are large, where they exceed the value of n$_e$, and are quoted as 100\%.}
\tnote[3]{`---' indicates electron densities $\lesssim$5 cm$^{-3}$.}
}
{\toprule[1.5pt]
Object		&	log(N/O)\tmark[1]	&	log(NII/OII)  		&	S~{\sc ii} line ratio\tmark[2]	&	n$_e$(cm$^{-3}$)\tmark[3] \ML
J0005-28 		& -1.683$\pm$0.031 & -1.641$\pm$0.029 & 1.360$\pm$0.092  & 57.5$\pm$18.7 \\
J1152-02A 	& -1.810$\pm$0.033 & -1.618$\pm$0.033 & 1.385$\pm$0.103 & 40.5$\pm$36.9 \\
J1152-02B 	& -1.880$\pm$0.037 & -1.715$\pm$0.036 & 1.404$\pm$0.099 & 24.0$\pm$23.0 \\
J1225-06s2 	& -1.822$\pm$0.054 & -1.663$\pm$0.054 & 1.432$\pm$0.158 & --- \\
J1328+02 	& -1.581$\pm$0.052 & -1.462$\pm$0.045 & 1.540$\pm$0.151 & --- \\
J1403-27 		& -1.697$\pm$0.035 & -1.574$\pm$0.032 & 1.408$\pm$0.106 & 20.5$\pm$100\% \\
J1609-04b2 	& -1.844$\pm$0.047 & -1.609$\pm$0.040 & 1.420$\pm$0.121 & 17.6$\pm$100\% \\
J1609-04b5 	& -1.630$\pm$0.051 & -1.544$\pm$0.047 & 1.455$\pm$0.141 & --- \\
J2039-63A 	& -1.667$\pm$0.044 & -1.562$\pm$0.040 & 1.344$\pm$0.114 & 73.2$\pm$100\% \\
J2039-63B 	& -1.738$\pm$0.067 & -1.499$\pm$0.067 & 1.501$\pm$0.224 & --- \\
J2234-04B 	& -1.792$\pm$0.069 & -1.554$\pm$0.069 & 1.389$\pm$0.240 & 35.5$\pm$100\% \\
J2242-06 		& -1.813$\pm$0.038 & -1.560$\pm$0.038 & 1.497$\pm$0.176 & --- \\
J2254-26 		& -1.612$\pm$0.033 & -1.451$\pm$0.032 & 1.317$\pm$0.095 & 96.8$\pm$34.0 \\
J2311-42A 	& -1.605$\pm$0.060 & -1.463$\pm$0.052 & 1.473$\pm$0.160 & --- \\
J2311-42B 	& -1.637$\pm$0.046 & -1.490$\pm$0.041 & 1.456$\pm$0.132 & --- \\
J2349-22 		& -1.702$\pm$0.059 & -1.628$\pm$0.051 & 1.457$\pm$0.150 & ---\\
\addlinespace[5pt]\bottomrule[1.5pt]\addlinespace[5pt]}

\FloatBarrier

\FloatBarrier 

\subsection{T$_e$ : Oxygen gas-phase abundance}

\begin{figure}[htbp]
\centering
\includegraphics[width=0.7\hsize]{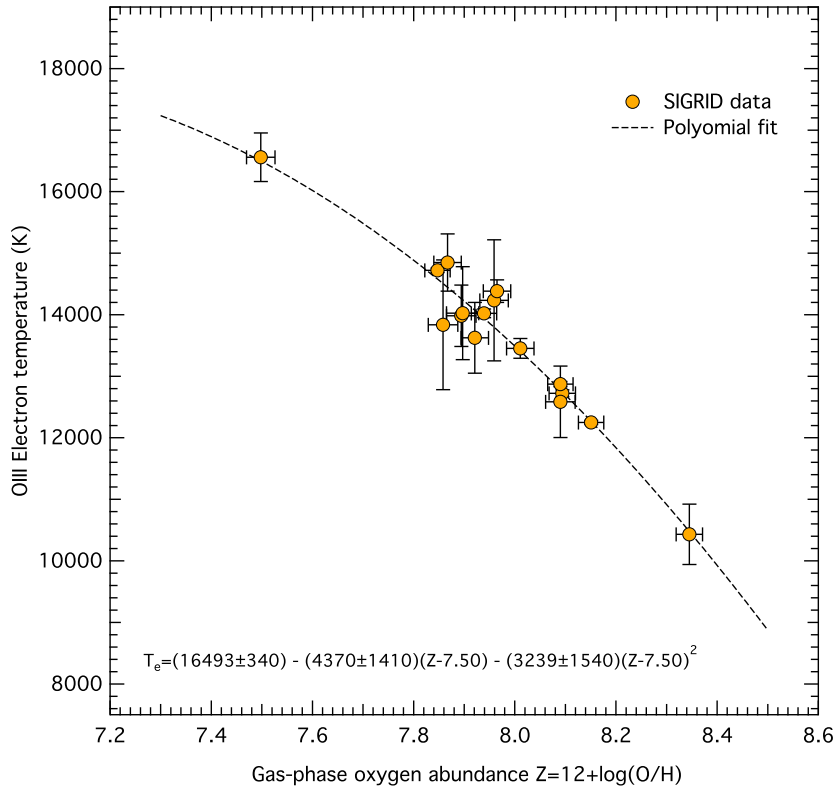}
\caption{Electron temperature, T$_e$, versus gas-phase oxygen abundance for the SIGRID objects, calculated using Equations \ref{eq2} and \ref{eq5}.}\label{fig_12}
\end{figure}

Figure \ref{fig_12} shows electron temperature, plotted versus gas-phase oxygen abundance, Z, from Table \ref{tx}.  Z (=12+log(O/H)) is derived using the formulae in Equations \ref{eq2} and \ref{eq5}. The quadratic fit to the data with 66\% confidence errors is:
\begin{equation}\label{teeq}
T_e = -(0.3239\pm0.1540)(Z-7.50)^2 -(0.4370\pm0.1410)( Z-7.50) +(1.6493\pm0.0340)) \times 10^4
\end{equation} 
While Figure \ref{fig_12} may be used to calculate the total gas-phase oxygen abundance from the [O~{\sc iii}] electron temperature, it applies only to the data presented here, and we will provide a substantially more accurate fit to the model-derived curve in paper 2.

\FloatBarrier
\section{Discussion}

\subsection{Mass--Metallicity}

\begin{figure}[h]
\includegraphics[width=\textwidth]{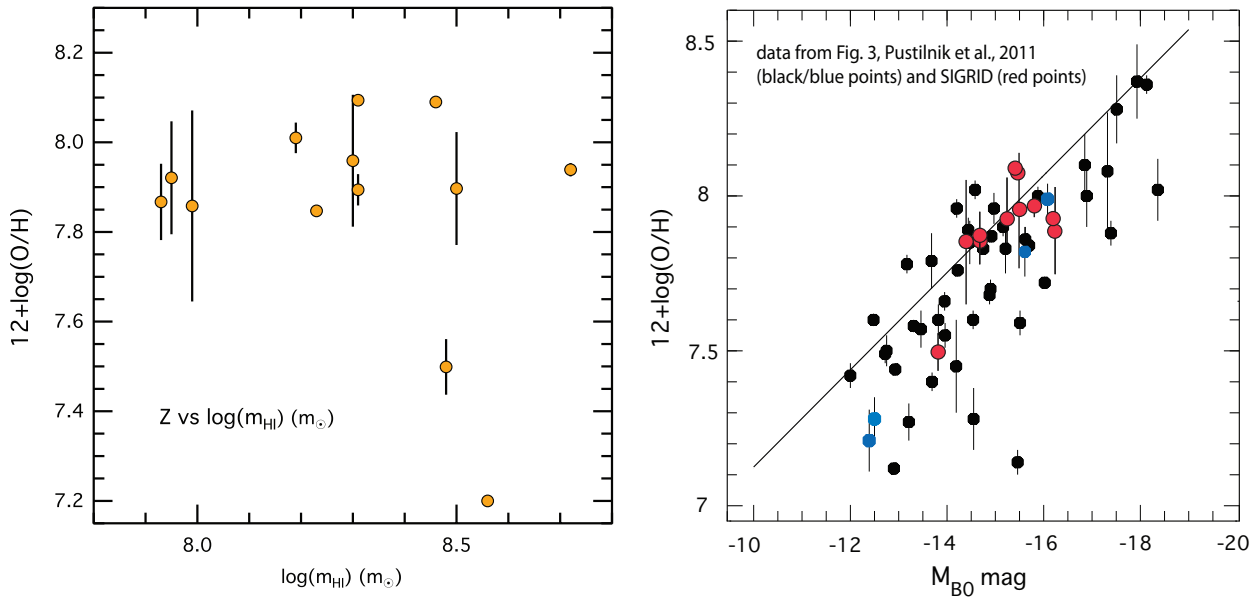}
\caption{Left Panel: Oxygen abundance versus neutral hydrogen mass (from Table \ref{t1}). Right Panel: Oxygen abundance versus absolute B-band magnitude, comparing data from Figure 3 from \citet{2011AstBu..66..255P} with SIGRID data. Eleven SIGRID objects have measured B-band magnitudes.  The four blue points are those from the Pustilnik sample which meet the luminosity, galaxy type, and and isolation selection criteria for SIGRID. The straight line fit is for field dI galaxies from \cite{2003AJ....125..146L}. The B-band magnitudes for the SIGRID sample have been taken from the NASA Extragalactic Database (NED).}\label{fig_13}
\end{figure}

Mass (or luminosity) versus metallicity behavior is one of the important evolutionary diagnostics for galaxies. It has been extensively mapped for larger galaxies \citep[e.g.,][]{2004ApJ...613..898T}, but it is less well known for dwarf galaxies. It has been studied by several authors \citep{2006ApJ...647..970L, 2011AstBu..66..255P,2012ApJ...754...98B,2013ApJ...765..140A}.  Exploring it was one of the initial motivations for the SIGRID sample \citep{2011AJ....142...83N}. Figure \ref{fig_13} shows the gas-phase oxygen abundance versus neutral hydrogen mass (left panel, data from \cite{2006ApJS..165..307M} and Table \ref{t1}) and gas-phase oxygen abundance versus absolute B-band magnitude (right panel).

There is no clear trend in the first graph, suggesting the neutral hydrogen mass is not strongly correlated with metallicity, at least in this sample. In the right panel, we compare the SIGRID data against data from the \cite{2011AstBu..66..255P} survey of galaxies in the Lynx-Cancer void.  The SIGRID data are consistent with the Pustilnik et al. data, and both samples are selected for isolation.  The four blue points in the right panel are objects from Pustilnik et al., but which meet the more stringent selection criteria for the SIGRID sample, for luminosity, galaxy type and isolation (distance from nearest neighbor). The trend line is from \cite{2003AJ....125..146L} for field dI galaxies, but which were not otherwise selected for isolation. Both the Pustilnik and SIGRID data tend to fall below the line, indicating that more isolated objects have slightly lower metallicities than similar objects in more congested regions, as suggested by \cite{2011AstBu..66..255P}.

There is insufficient data in our observations to confirm the increasing spread of metallicity values at low mass, as implied by \citet[][Figure 6]{2004ApJ...613..898T}. However, the log(N/O) data (Figure \ref{fig_10}) are consistent with such a spread.

\FloatBarrier 

\subsection{Comparison of metallicity methods}

In this work, we have calculated the gas-phase oxygen abundance using the electron temperature direct method routine developed here, and the diagnostic grids. Table \ref{tx} compares the electron temperature abundances using the methods described here with those using the iterative method from \cite{2006A&A...448..955I}, with the same input temperatures; and the most reliable values from the diagnostic grids, those using the log(N~{\sc ii}/S~{\sc ii}) and log(N~{\sc ii}/O~{\sc ii})diagnostics.   The values derived using the two T$_e$ methods are similar, suggesting that the method developed here is reliable.  See also a discussion of this in the forthcoming paper 2.

It is interesting to note that the diagnostic grid abundances, with two notable exceptions, are consistently a little higher than the direct method values, consistent with the findings of \cite{2012MNRAS.426.2630L}.  The average value of the difference $\delta$Z is 0.104$\pm$0.171.  The complete explanation of this difference is unclear, but in part it can be explained by the nature of the diagnostic grids derived from the Mappings photoionization modelling code.  In the grids, we have assumed a dust depletion of 1.0 solar \citep{2013ApJS..208...10D}, to account for the elements locked up in dust grains.  This leads to an overestimate of 0.07 dex in the abundance values derived from the diagnostic grids, compared to the gas-phase-only oxygen abundances from the direct method, explaining about half of the discrepancy.

It is useful to examine two of the ``outliers'' in Table \ref{tx}, where the diagnostic grid oxygen abundances differ substantially from the electron temperature oxygen abundances. J1225-06s2  has a very low oxygen abundance, $\sim$7.45, from the direct method, and $\sim$7.9 from the grids.  This could be explained if there is  more N~{\sc ii} than implied by the Mappings models parameters, although this is not obvious from Table \ref{ty}.  There may also be increased nitrogen due to enrichment by WN stars, as in the case of HS 0837+4717 \citep{2004A&A...419..469P,2011A&A...532A.141P}. The oxygen abundance discrepancy for J1609-04(2) is very likely a result of uncertainty in the flux of the [O~{\sc iii}] 4363 \AA~ auroral line, which is weak in this object.

The diagnostic grids themselves (Table \ref{t3}) are somewhat discrepant in the values yielded for oxygen abundances.  In particular, two trends are clear.  First, diagnostics involving log(N~{\sc ii}/S~{\sc ii}) are particularly consistent, and the closest to the oxygen abundances derived using the direct method. Diagnostics involving [O~{\sc ii}] fluxes are nearly as consistent.  This concordance and consistency lead us to believe that these diagnostics are the most reliable, and we have used the means of the log(N~{\sc ii}/S~{\sc ii}) and log(N~{\sc ii}/O~{\sc ii})diagnostics in Table \ref{tx}.  Second, diagnostics involving log(N~{\sc ii}/H$\alpha$) give somewhat higher oxygen abundances than both the other diagnostics and the direct method values.  The source of these discrepancies is unclear, but may be related to the abundance fit for nitrogen used in Mappings.  They do not materially affect the results reported here, provided we rely on the log(N~{\sc ii}/S~{\sc ii}) and log(N~{\sc ii}/O~{\sc ii}) diagnostics and direct method oxygen abundances.

\subsection{Further analysis and investigations}

In the second paper examining these observations, we will explore the effect of three-dimensional diagnostic charts. These use three independent diagnostic ratios plotted and explored in three dimensions, whose purpose is to investigate whether the observations lie on a diagnostic plane, along the lines of Vogt at al. (2014, in prep.).   We will examine the effects of using diagnostic grids calculated using higher electron densities.  We will investigate the effects of optically thin \HII regions, and show that they can have considerable effects on the diagnostics, and that there is evidence of optical thinness in some of the observed objects. We will re-examine the electron temperature versus oxygen abundance plots, for both these observations and for 124 SDSS objects from \cite{2006A&A...448..955I}. Using the Mappings photoionization modelling code, we will demonstrate that with reasonable assumptions about the star clusters exciting \HII regions, there is an effective upper limit to the temperature that can be reached, even in optically thin regions.  The implications appear to be that some of the high temperatures reported in low metallicity \HII regions may be somewhat in error.  We will demonstrate the effect of taking into account the additional contribution to total oxygen abundance of the oxygen in dust grains.  We will also suggest that the apparent spread in metallicities at the low end of the mass-metallicity relation are due to stochastic effects in stellar mass distributions in the small star clusters exciting \HII regions in small irregular galaxies.

\section{Conclusions}

In this paper we have presented the results of observations of seventeen \HII regions in thirteen small isolated dwarf irregular galaxies, most from the SIGRID sample, all but one exhibiting the [O~{\sc iii}] auroral line.  All have measured oxygen abundances $<$8.2 ($<$0.3 Z$\odot$), one has an apparent abundance of 7.44 and another very low metallicity object with Z$\sim$7.2. We have derived a method for calculating total gas-phase oxygen abundances using only the optical spectra between 3700 and 7000 \AA. This method gives very similar results to previous empirical fit methods. From an analysis of abundances and ionization coefficients using the diagnostic grids developed by \cite{2013ApJS..208...10D}, we find the direct method oxygen abundances are consistently within 0.07 dex of the strong line diagnostic results, making allowance for the oxygen locked up in dust grains.  From the line ratio of the two red [S~{\sc ii}] lines we find that the electron densities occurring in the objects observed are between $\sim$5 and 100 cm$^{-3}$. The nitrogen abundance, as expressed in log(N/O), continues the trend evident in \cite{1998AJ....116.2805V}, but from this sample we find no clear evidence for a nitrogen floor.  There is increased scatter at lower oxygen abundances, and some evidence for a bifurcation in the trend, possibly due to the presence of WN stars in some of the \HII regions. The slope of the luminosity--metallicity relation for these observations is very close to that for void galaxies in \cite{2011AstBu..66..255P}.  The spectra from an apparently very low metallicity galaxy, J1118-17s2, show no nitrogen lines: we intend to undertake follow up observations on this galaxy to estimate the metallicity more accurately.

\begin{acknowledgments}
Mike Dopita acknowledges the support of the Australian Research Council (ARC) through Discovery  project DP0984657. This work was funded in part by the Deanship of Scientific Research (DSR), King Abdulaziz University, under grant No. (5-130/1433 HiCi). The authors acknowledge this financial support from KAU.
\end{acknowledgments}

%\bibliography{/Users/...../Documents/bibdesk/references}

\bibliographystyle{aa}

\newpage
\section{Appendix: Emission line flux error estimation}

This appendix describes the methods used to estimate emission line flux uncertainties for spectra extracted from WiFeS data cubes of objects in the SIGRID sample. The data reduction process is described in detail in \cite{2013Ap&SS.tmp..406C}, and in this paper in Section 2.3. Briefly, the steps where noise is involved or systematic errors are incurred are bias subtraction, flat fielding, cosmic ray removal, sky-line subtraction using nod-and-shuffle, and standard flux star calibration. The principle sources of uncertainty are the CCD detector and amplifier readout noise, and the amplification of this noise through the data reduction chain; the effects of cosmic rays and sky lines, and their removal (partial or complete); the calibration of the emission line fluxes using standard star flux data; the de-reddening process; and the measurement of the line fluxes from the flux-calibrated spectra.

As the IFU data frame is convolved into a data cube in the pipeline, the process of error calculation is  more complex than for single slit or echelle spectroscopy.  For the spectral noise uncertainties, there are two approaches we could take. One is to estimate the errors accumulating from each step, such as described for echelle spectroscopy by \cite{1994ApJ...431..172S}. The other approach, used here, is to  measure the statistical noise from spectra extracted from the reduced data cube, and to estimate the systematic errors arising from the de-reddening and flux calibration, which are independent of the statistical noise. Unlike single slit spectra, with IFU data cubes, we are able to select the entire area of the \HII region from which to extract the spectrum, and exclude the majority of the galaxy stellar background, resulting in better signal-to-noise. Note that the statistical noise varies with the size of the sampled spaxel area, due to averaging. For the objects in this study, sampling using a 6 arc sec diameter circular spatial area maximises the amount of flux from the \HII region and minimises both the statistical noise, though averaging, and the stellar continuum from the area outside the \HII region.

In every case, the galaxies were so faint that the stellar extent was at best barely detectable. However, images from the DSS survey and from the SINGG data \citep[as illustrated in][]{2011AJ....142...83N} suggest the individual galaxies are less than or approximately equal to the FOV of the WiFeS spectrograph, 25$\times$38 arc sec. The benefit of the IFU is that the sample was centered on the \HII region, and excluded virtually all areas of the galaxies without \HII emission. 

Line fluxes and noise were measured from the extracted spectra using IRAF/splot.  The standard splot `k-k' method was used to fit a gaussian to each emission line, to measure the equivalent width (where possible), the gaussian full width at half maximum, and the integrated flux. Noise was measured on both sides of the emission line using the splot `m-m' method. These results were checked using the deblend `d-d' method, but using a single line, which automatically generates values of the same parameters. Particular care was taken to account for any stellar absorption features underlying the Balmer emission lines, although in all cases, this was minor or absent, due to low stellar continuum.  In fact, the stellar continuum was extremely faint, with the exception of the object J0005-28 (see Figure \ref{fig_7}, displayed on a log-intensity scale).  Test sample sizes showed that all the detectable H$\alpha$ and [O~{\sc iii}] in each \HII region lay within the sample aperture. The observed fluxes mostly peak at or less than a radius of 2.5 arc sec, except where there are closely adjacent \HII regions (e.g., J1609-04). For these, limiting the sample size to 6 arc sec diameter avoided sampling a different \HII region. Ideally, single spaxel-based analysis would be preferable to multi-spaxel sampling, but these objects are so faint that the resultant noise is prohibitive.

Detector noise is added to the data frame during bias subtraction and flat fielding, as the bias and flat field frames used also incur readout noise.  The sky subtraction process using the nod and shuffle process or the sky frame method adds additional noise during the subtraction process. Nod-and-shuffle sky subtraction was used for all \HII region observations, with sub-exposure times chose to be shorter than the shortest observed fluctuation in the OH airglow lines \citep{2000GeoRL..27...41F}. The removal of the critical OH lines is effectively complete in all observations. The [O~{\sc i}] airglow lines are at wavelengths that did not interfere with any of the observed \HII region spectral lines.

\begin{figure}[htpb]\label{cosray}
\centering
\includegraphics[scale=0.5]{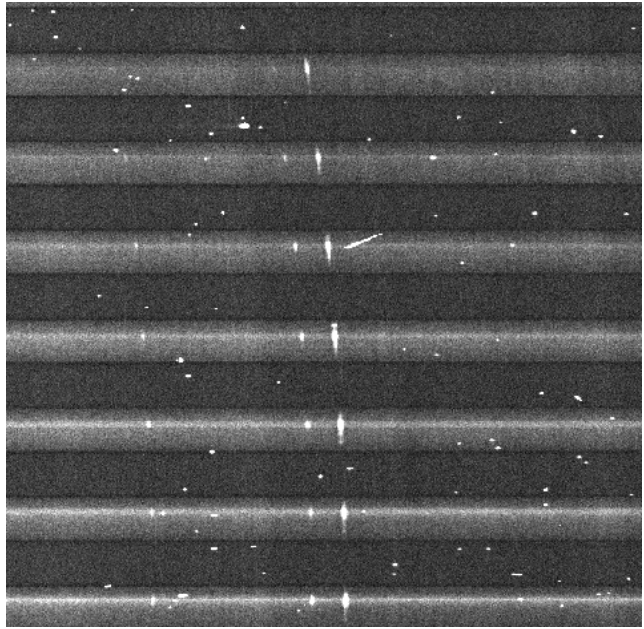}
\caption{Section of a single WiFeS IFU raw data frame for  J0005-28 with cosmic ray artefacts, centered on the H$\gamma$ line, showing sections of 7 slitlets. Each frame was inspected to check for cosmic ray contamination of key lines. Note that, due to the optical paths in the WiFeS IFU blue camera, the right side of the image corresponds to shorter wavelengths.}\label{figa0}
\end{figure}

Cosmic ray removal is reasonably efficient, using the Laplacian kernel technique described by \cite{2001PASP..113.1420V}. The process is not perfect, but virtually all the remaining cosmic ray artefacts are removed using the \textsf{imcombine} process.  In isolated cases, separate cosmic rays occur on all object data frames at the same location, and this can lead to erroneous results, but this can be detected by the labor intensive process of inspecting all the lines on all the IFU slitlets (25) on all the data frames (usually 3).  Figure \ref{cosray} shows part of a raw WiFeS blue data frame including segments of 7 slitlets (of a total 25), centered on the H$\gamma$ and [O~{\sc iii}] $\lambda$ 4363  auroral line, for the galaxy J0005-28.  The auroral line is very prominent in this frame, to the left of the H$\gamma$ line.

%\FloatBarrier

In addition to the intrinsic statistical noise amplified through the data reduction pipeline, when measuring the emission line fluxes, it is necessary to take into account any broad absorption lines in the stellar continuum, on which the nebular spectra are superimposed. There are three approaches here.  The first is to correct for an assumed 2 \AA~ EW absorption in each line, as described in  \cite{1994ApJ...431..172S}.   The second is to use an automated method such as the LZIFU IDL program developed by several workers at the University of Hawaii, which fits model stellar continua to observed spectra and then calculates the emission line fluxes (a paper on this application is planned).  The third approach, which we use here, works better when the stellar continuum is weak, as with the objects reported here. It involves manual fitting of  gaussian profiles to the emission lines using standard IRAF/splot methods. The technique described by \cite{2013ApJ...775..128B}  is very similar in detail to the method used here.

The errors arising from the de-reddening process are due to uncertainties in the nature and amount of dust between the nebular emission and the observer.  In the case of the SIGRID  objects considered here, all are further than 10$^\circ$ from the galactic equator, to avoid significant reddening by Milky Way dust. We calculated the de-reddening using two independent methods and used the differences between the results as an estimate of the de-reddening errors. We used the dust reddening formulae from \cite{1989ApJ...345..245C} with A$_V$=3.1, adjusting the de-reddening to set the resultant Balmer H$\alpha$/H$\beta$  flux ratios to the \cite{1995MNRAS.272...41S} Case B Balmer ratios for the calculated [O~{\sc iii}] electron temperature.  We used the ratios of H$\gamma$ and H$\delta$ to H$\beta$ as confirmation.  To confirm these results, we employed the dust models from \cite{2005ApJ...619..340F}, using a relative extinction curve with R$^A_V$=4.3, where R$^A_V$ = A$_V$/(E$_{B-V}$) and A$_V$ is the V-band extinction.  This is discussed in more detail in \citet[Appendix 1]{2013ApJ...768..151V}. We used an initial Balmer decrement ratio of 2.82 for H$\alpha$/H$\beta$, corresponding to an electron temperature of 12 500K, adjusted the electron temperature using the direct method derived from the [O~{\sc iii}] line ratios, then adjusted the apparent Balmer ratios by varying the value of A$_V$ for the best fit to the H$\gamma$/H$\beta$ ratio, using the ratio H$\delta$/H$\beta$ as a check, again fitting to the Storey and Hummer Case B Balmer ratios. 

The de-reddened flux values reported in Table \ref{t2}  are those using the Cardelli method. In all cases, the two approaches gave similar results: The average difference between the two methods for the important diagnostic lines varies between 0.1\% and 0.7\% . As a consequence, we have adopted a figure of 1\% for the de-reddening error.  In only one case, J2234-04, object A, did the de-reddening fail to provide a plausible result, and this has been excluded from the results reported here.  It appears likely that two or more incompletely removed sets of cosmic ray artefacts were the cause of the problem, in this particular case.

Flux calibration errors depend on how well one can fit flux-calibrated vales to the standard stars. The standard stars used were taken from \cite{1999PASP..111.1426B}, using Bessell's recalibration of Hamuy's Southern Spectrophotometric Standards \cite{1992PASP..104..533H} . Calibrating to these standards is likely to be more accurate than the older Oke standards \citep{1990AJ.....99.1621O}, but we have retained an estimated 2\% error for calibrating the flux values, as per \cite{2013ApJ...775..128B}.

We have recently corrected a problem with the measured [O~{\sc ii}] line fluxes. The source of the problem is a sharp absorption edge at 3850\AA, due to the adhesive used in the beam splitter, and the lack of any significant output below about 3900\AA~from the lamps used for flat exposures to identify and correct this problem. The lamps are now being replaced, and future measurements using WiFeS will no longer require this compensation. The problem was understood in the testing phase of the construction of WiFeS, and was identified in the observations from poor matches to the diagnostic grids that involve [O~{\sc ii}] in the ratios. To compensate requires boosting the [O~{\sc ii}] flux by a factor of $\times$1.5. It affects only the [O~{\sc ii}] $\lambda\lambda$ 3726,9 lines. It does not make a substantial difference to the calculated oxygen abundances or other results.

\end{document}